\documentclass[aps,footinbib,prl,superscriptaddress,reprint]{revtex4-1}
\usepackage{graphicx}
\usepackage{amssymb}
\usepackage{hyperref}
\usepackage{amsmath}
\usepackage{slashed}
\usepackage{lipsum}
\usepackage[dvipsnames]{xcolor}
\usepackage{xcolor}   
\usepackage{xspace} 
\usepackage{bm}
\usepackage{xfrac}
\usepackage{lineno}
\usepackage[normalem]{ulem}
\usepackage[toc,page]{appendix}
\usepackage[capitalize]{cleveref}
\usepackage{nicefrac}

\usepackage[T1]{fontenc}      
\usepackage{ae} 
\usepackage{grffile}

\newcommand{\im}{\ensuremath{\mathrm{i}}}

\newcommand{\RM}[1]{\MakeUppercase{\romannumeral #1{}}}
\newcommand*\rfrac[2]{{}^{#1}\!/_{#2}}
\DeclareMathOperator{\sign}{sgn}

\RequirePackage{fix-cm} 
\DeclareMathSizes{9.5}{9.5}{7.0}{5}

\begin{document}

\title {Survival of the quantum anomalous Hall effect in orbital magnetic fields as a consequence of the parity anomaly}
\author{Jan~B\"ottcher}
\thanks{These two authors contributed equally to this work.}
\affiliation{Institut f\"ur theoretische Physik (TP4) and W\"urzburg-Dresden Cluster of Excellence ct.qmat, Universit\"at W\"urzburg, 97074 W\"urzburg, Germany}
\author{Christian~Tutschku}
\thanks{These two authors contributed equally to this work.}
\affiliation{Institut f\"ur theoretische Physik (TP4) and W\"urzburg-Dresden Cluster of Excellence ct.qmat, Universit\"at W\"urzburg, 97074 W\"urzburg, Germany}
\author{Laurens~W.~Molenkamp}
\affiliation{Physikalisches Institut (EP3), Universit\"at W\"urzburg, Am Hubland, 97074 W\"urzburg, Germany}
\author{E.~M.~Hankiewicz} 
\email{Ewelina.Hankiewicz@physik.uni-wuerzburg.de}
\affiliation{Institut f\"ur theoretische Physik (TP4) and W\"urzburg-Dresden Cluster of Excellence ct.qmat, Universit\"at W\"urzburg, 97074 W\"urzburg, Germany}

\begin{abstract}
Recent experimental progress in condensed matter physics enables the observation of signatures of the parity anomaly in two-dimensional Dirac-like materials. Using effective field theories and analyzing band structures in external out-of-plane magnetic fields (orbital fields), we show that topological properties of quantum anomalous Hall (QAH) insulators are related to the parity anomaly. We demonstrate that the QAH phase survives in orbital fields, violates the Onsager relation, and can be therefore distinguished from a quantum Hall (QH) phase. As a fingerprint of the QAH phase in increasing orbital fields, we predict a transition from a quantized Hall plateau with $\sigma_\mathrm{xy}= -\mathrm{e}^2/\mathrm{h}$ to a not perfectly quantized plateau, caused by scattering processes between counterpropagating QH and QAH edge states. This transition can be  especially important in paramagnetic QAH insulators, such as (Hg,Mn)Te/CdTe quantum wells, in which exchange interaction and orbital fields compete. 
\end{abstract}

\maketitle

\textit{Introduction.}
Condensed matter analogs of the Dirac equation have opened new directions to study quantum anomalies in the solid-state laboratory \cite{Semenoff84,Fradkin86,Haldane88,Huang15,Li15,Zhang16,Yan17,Gooth17}. An anomaly occurs, when a symmetry of a classical theory cannot be maintained in the associated quantum theory  \cite{Fujikawa80,Bertlmann96,Nakahara2003}. For instance, in massless, (2+1)D quantum electrodynamics, parity symmetry is broken during regularization if one insists on gauge invariance (parity anomaly)  \cite{Niemi83,Niemi84,Jackiw84,Niemi85,Redlich84,Boyanovsky86,Schakel91,Burnell13}. As a consequence, a Chern-Simons (CS) term is induced even in the absence of a magnetic field \cite{Redlich84,Haldane88}.


In condensed matter physics, an analogous system is a Chern/quantum anomalous Hall (QAH) insulator which describes a single Dirac fermion with a momentum dependent mass or, equivalently, half of the Bernevig-Hughes-Zhang (BHZ) model \cite{Bernevig06,Liu08}. 
In our work, we examine hallmarks of the parity anomaly in two-dimensional QAH insulators subjected to an external out-of-plane magnetic field (orbital field). In particular, we demonstrate that the parity anomaly enables us to distinguish the QAH from a quantum Hall (QH) phase. 
This is due to the fact, that although both phases are described by the same topological invariant, the Chern number \cite{Schnyder08}, their physical origin is very different: QH phases are induced by an orbital field, whereas the  QAH phase results from an inverted band structure \footnote{We take the spin-up block of the BHZ model with a non-trivial Chern number $\mathcal{C} = -1$ at $H=0$.}. Here, inverted means that the ordinary conduction band is below the ordinary valence band. 
A QAH insulator is characterized by a quantized Hall conductivity $\sigma_\mathrm{xy}=\mathcal{C} \, \mathrm{e}^2 / \mathrm{h}$ with $\mathcal{C}=\left[\mathrm{sgn}(M)+\mathrm{sgn}(B)\right]/2$ \cite{Lu10}, where $2M$ is the bulk band gap (Dirac mass gap) and $B$ is related to the effective mass. In our work, we reveal that this characteristic quantity persists in orbital fields $H$ with

\begin{equation} \label{eq:Intro}
\mathcal{C}(H)=\left[ \mathrm{sgn} \left(M-B/l_H^2 \right) + \mathrm{sgn} \left( B \right) \right]/2 \ ,
\end{equation}
where $l_H=\sqrt{\hbar/\vert \mathrm{e} H\vert}$. Equation~\eqref{eq:Intro} shows that $H$ counteracts the intrinsic band inversion until it eventually overcomes the Dirac mass gap at  $M = B/l_{H_\mathrm{crit}}^2$. 
Moreover, it illustrates a violation of the  Onsager relation which would require that $\sigma_\mathrm{xy}(-H)=-\sigma_\mathrm{xy}(H)$. This is a hallmark of the parity anomaly in magnetic fields.
In contrast, a conventional QH phase fulfills the Onsager relation as $\sigma_\mathrm{xy} \propto \sign (\mathrm{e}H)$.


As a signature of the parity anomaly, the survival of the QAH phase induces a unique type of charge pumping. 
Increasing the orbital field generates a charge flow from the edges (charge depletion) into the bulk (charge accumulation), starting at $H \neq 0$.  Moreover, as a function of $H$, the QAH edge states are pushed into the bulk valence band, leading to the coexistence of  counterpropagating QH and QAH edge states. If  disorder is present, these states are not protected from backscattering. We predict, that these two effects give rise to a system size dependent transition from $\sigma_\mathrm{xy}=-\mathrm{e}^2/\mathrm{h}$ to a not perfectly quantized Hall plateau. The average value  of  this plateau depends on details of the scattering mechanisms. Such a transition should be observable  in (Hg,Mn)Te quantum wells \cite{Budewitz17} or in Bi-based QAH insulators \cite{Chang13,Checkelsky14,Moodera15,Bestwick15}.

\begin{figure}[!t]
\centering
\includegraphics[width=0.98\columnwidth]{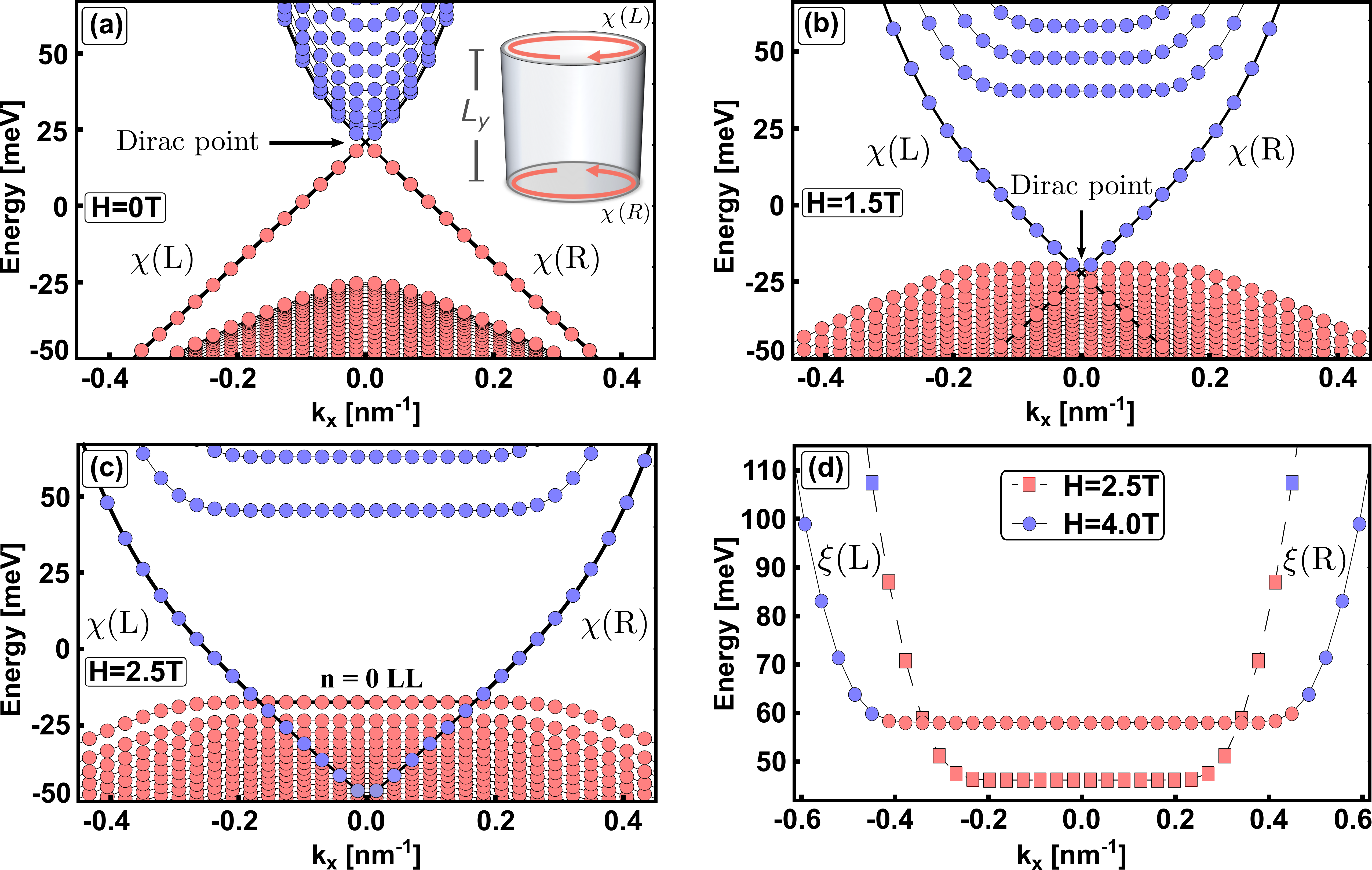}
\caption{\label{fig:QAHEvolution} Band structure of a QAH insulator in orbital fields $H$ (black lines) for ${M=-25\, \text{meV}}$, $B=-1075\, \text{meVnm}^2$, $D=-900\, \text{meVnm}^2$, and $A=365\,\text{meVnm}$. $\chi(\mathrm{L/R})$ and $\xi(\mathrm{L/R})$ depict QAH and QH edge states at the left and  right boundary.  (a) Spectrum for $H\!=\!0$ at half filling with chiral QAH edge states traversing the bulk gap. The inset depicts the sample geometry. (a)-(c)  Evolution of the spectrum and its filling  with \mbox{increasing $H$}, where empty/filled states are marked in blue/red. (d) Analogous analysis for an initially filled conduction band LL.}
\vspace{-0.4cm}
\end{figure}

\textit{Model.}
We start with a Chern/QAH insulator described by a single, non-trivial block of the BHZ model:
\begin{align}
\! \! \! \mathcal{H} ( \mathbf{k} )=\left(M \! -B k^2\right)\sigma_\mathrm{z}-D k^2 \sigma_{0}+A\left(k_\mathrm{x} \sigma_\mathrm{x}-k_\mathrm{y} \sigma_\mathrm{y} \right),
\label{eq:ChernHamilton}
\end{align}
where $k^2 = k_\mathrm{x}^2+k_\mathrm{y}^2$, $\sigma_i$ are the Pauli matrices, $A$ mixes both (pseudo)spin-components, $D$ introduces a particle-hole asymmetry, and  $B$, as well as $M$ were defined before \cite{Bernevig06}. The spectrum is obtained numerically by mapping the Hamiltonian on a stripe geometry with finite length $L_\mathrm{y}$ in the $\mathbf{e}_\mathrm{y}$-direction (hard wall boundary conditions) and periodic boundary conditions along the $\mathbf{e}_\mathrm{x}$-direction \cite{Scharf12}. In Fig.~\ref{fig:QAHEvolution}(a), the band structure with $\mathcal{C}=-1$ is displayed, with chiral QAH edge states traversing the Dirac mass gap.  Since $D\neq0$, the Dirac point lies close to the conduction band edge \cite{Zhou08}.

Next, we implement an orbital field $\mathbf{H}=H \mathbf{e}_\mathrm{z}$ in the Landau gauge $\mathbf{A} = -y H \mathbf{e}_\mathrm{x}$  [Figs.~\ref{fig:QAHEvolution}(b)-(c)]. This has two main effects: First, bulk subbands evolve into Landau levels (LLs) for $l_H \ll L_\mathrm{y}$. All LLs with ${n\!\neq\!0}$ come in pairs of energy $E_n^\pm$, except for the single ${n\!=\!0}$ LL with energy $E_\mathrm{0}$ \cite{Konig08}. This causes an asymmetry in the spectrum further discussed in App.~B.
Second, the orbital field gradually lowers the energy of the Dirac point so that it enters the valence band at $H=H_\mathrm{scat}$. 
The evolution of the Dirac point is determined by ${E_\mathrm{D}(H) \approx E_\mathrm{D}(0)-g_\mathrm{eff}  \mu_\mathrm{B} H}$,
where $g_\mathrm{eff} = \mathrm{m}_0 v_\mathrm{x} L_\mathrm{y} / \hbar$ \cite{Zhou08}. Here, $v_\mathrm{x}$ is the edge state velocity, $\mu_\mathrm{B}$ is the Bohr magneton, $\mathrm{m}_0$ is the electron mass, and $E_\mathrm{D}(0)$ is the Dirac point energy at $H = 0$.
Note that the QAH edge states survive (up to finite size gaps) even for large $H$ [Figs.~\ref{fig:QAHEvolution}(a,c)] since they are protected from hybridization with bulk states by their wave function localization.


\textit{Effective Action.}
To understand the survival of the QAH edge states, we derive the corresponding low energy effective bulk Lagrangian $\mathcal{L}_\mathrm{eff}^\mathrm{bulk}$ by computing  the  particle number in the continuum/bulk model \cite{Niemi85},
\begin{align*} 
\left\langle N \right\rangle_\mu \!= \!\frac{1}{2}\int\!\! \mathrm{d}\mathbf{x}\sum_{\alpha} \left\langle  \left[\psi^\dagger_\alpha(\mathbf{x}),\psi_\alpha(\mathbf{x})\right]\right\rangle_\mu \!= \left\langle N_\mathrm{0}\right\rangle_\mu\!-\!\frac{\eta_{_H}}{2}.
\end{align*}
Here,  $\langle \ldots \rangle_\mu$  denotes the expectation value with respect to the chemical potential $\mu$, $\psi(\mathbf{x})$ is a field operator, and $N_\mathrm{0}$ is the fermion number operator, counting the number of filled/empty states with respect to the charge neutrality point. The last term is the spectral asymmetry $\eta_{_H}$ \cite{Niemi85}, quantifying the difference in the number of positive and negative eigenvalues of our system. 
From Lorentz covariance, one can then determine the induced three current $j^\mu_\mathrm{ind}=\sigma_{\mathrm{xy}} \epsilon^{\mu\nu\rho}\partial_\nu a_\rho$ arising as a response to a small perturbing field $a_\mu$, applied on top of the underlying orbital field $H$.
Here, $j^0_\mathrm{ind}$ is the induced bulk charge density,  and $j^\mathrm{1,2}_\mathrm{ind}$ is the induced bulk current density in x- and y-direction, respectively. 
Since $j^\mu_\mathrm{ind}\!=\!\delta S_\mathrm{eff}^\mathrm{bulk} / \delta a_\mu$ with $S_\mathrm{eff}^\mathrm{bulk}=\int \! \mathrm{d}^3 x \, \mathcal{L}_\mathrm{eff}^\mathrm{bulk}$, we can compute the corresponding effective bulk Lagrangian which is one of the main results of our paper (further details are given in App.~B):
\begin{align} \label{sigmaxyges} 
\mathcal{L}_\mathrm{eff}^\mathrm{bulk} (\mu,H)= \frac{\sigma_\mathrm{xy}(\mu,H)}{2} \, \epsilon^{\mu \nu \rho} a_\mu \partial_\nu a_\rho  \,  ,
\end{align} 
where $\epsilon^{\mu \nu \rho}$ is the Levi-Civita symbol.
This is a topological CS term \cite{Deser82} with quantized Hall conductivity
\vspace{-0.1cm}
\begin{align} \nonumber
\! \!\sigma_{\mathrm{\scriptstyle{xy}}}    =&  \kappa_{_\mathrm{QAH}}  - \kappa_{_\mathrm{QH}}^{0}  \Theta  \left ( \vert \mu \! + \! D/l_H^2 \vert \! - \! \left \vert M \! - \!B/l_H^2  \right \vert  \right)   \\ 
& -  \sum_{\substack{s=\pm, \, n =1 }}^{\infty} \! \! s \kappa_{_\mathrm{QH}}   \Theta \left[ s (\mu \! - \! E_n^s ) \right] \ . 
\end{align}
According to their physical origin, we separated $\sigma_\mathrm{xy}$ into: \vspace{-0.1cm}
\setcounter{equation}{3}
\begin{subequations}  \label{allkappa}
\begin{align}
\! \! \kappa_{_\mathrm{QAH}}  & = \frac{\mathrm{e}^2}{2 \mathrm{h}}\left[ \mathrm{sgn}(M \! - \! B/l_H^2) + \mathrm{sgn}(B) \right] , \label{kappa1}\\ 
\kappa^{0}_{_\mathrm{QH}} & =  \frac{\mathrm{e}^2}{2 \mathrm{h}} \left[ \mathrm{sgn}( \mathrm{e} H) \, \mathrm{sgn}( \mu \! + \! D/l_H^2)  \nonumber \right. \\
 & \left. \hspace{2.2cm} + \,  \mathrm{sgn}(M \! - \! B /l_H^2) \right] , \label{kappa2}\\
\kappa_{_\mathrm{QH}} & =   \frac{\mathrm{e}^2}{ \mathrm{h}} \mathrm{sgn}(\mathrm{e} H) \, . \label{kappa3}
\end{align}
\end{subequations}
CS terms arise if parity and time-reversal symmetry are broken \cite{Deser82,Redlich84}. In our case, they are therefore induced by the mass terms $M$ and $B k^2$, as well as by the orbital field $H$ [App.~C]. 
We distinguish two types of CS terms: The first type, Eq.~\eqref{kappa1}, is defined by its exclusive relation to $M$ and $B k^2$, resulting in the violation of the Onsager relation. This term is a consequence of the parity anomaly at $H=0$, which requires that a single, parity invariant Chern insulator cannot exist in (2+1)D \cite{Redlich84}.  Its special origin is reflected by the fact that Eq.~\eqref{kappa1} is solely determined by the spectral asymmetry $\eta_{_H} \! = \! 2 \,n_0 \, \mathrm{sgn}(\mathrm{e}H)\kappa_{_\mathrm{QAH}} \mathrm{h} / \mathrm{e}^2$, where $n_0$ is the LL degeneracy. It is a property of the entire eigenvalue spectrum and, hence, does not come along with a Heaviside function.
The second type of CS terms, Eqs.~\eqref{kappa2} and \eqref{kappa3}, describes conventional QH physics, generated by an orbital field, as indicated by their $\mathrm{sgn}(\mathrm{e} H)$-dependence. 
In contrast to the first type, each of these CS terms is related to a single LL, reflected by the Heaviside functions. They can only contribute to the Hall conductivity if $|\mu  + \! D/l_H^2|\! > \! |M \! - \! B/l_H^2|$.

In order to derive the corresponding edge theories, we have to add a new degree of freedom to $\mathcal{L}_\mathrm{eff}^\mathrm{bulk}$. This can be inferred from the fact that any CS term changes by a total derivative under a local gauge transformation, $\mathcal{L}_\mathrm{eff}^\mathrm{bulk}\rightarrow \mathcal{L}_\mathrm{eff}^\mathrm{bulk}+ \delta \mathcal{L}_\mathrm{eff}^\mathrm{bulk}$,  causing a violation of charge conservation, $\partial_\mu j^\mu_\mathrm{ind} \neq 0 \  \vert_{_{\partial \Omega}}$, at the boundary $\partial\Omega$ \cite{Deser82,Wen91}.
To cancel this U(1)-anomaly, we must enlarge our description by an effective edge Lagrangian $\mathcal{L}_\mathrm{eff}^{\partial \Omega}$, which restores gauge invariance via anomaly cancellation between edge and bulk (Callan-Harvey mechanism) \cite{Chandrasekharan84,Callan85,Wen91,Nakai17}:
\begin{align} \label{anomalycancelation}
\hspace{-5cm} & \partial_\mu j^\mu_\mathrm{tot} = \partial_\mu \left( j^\mu_\mathrm{ind} + j^\mu_\mathrm{L} +j^\mu_\mathrm{R} \right)=0
\\
\Rightarrow  \  & \partial_\mu j^\mu_{_\mathrm{L/R}} \! = \dfrac{\sigma_\mathrm{xy}}{2} \,  \delta\!\left(y\!-\!y_{_\mathrm{L/R}} \right)  \epsilon^{2 \nu \lambda} \partial_\nu a_\lambda = \!- \partial_\mu j^\mu_\mathrm{ind} \, , \nonumber
\end{align}
where $j^\mu_{_\mathrm{L/R}}$  symbolizes induced currents at the left/right edge of the stripe geometry. This procedure is the field-theoretical analog to the bulk-boundary correspondence \cite{Fradkin2013}.
Equation~\eqref{anomalycancelation} implies that an orbital field induces charge accumulation  in the bulk which is compensated by a charge depletion at the edges (fixed total charge) \cite{Boyanovsky86,Stone87,Maeda96}. The amount of induced bulk charge is given by $j^0_\mathrm{ind}=\sigma_\mathrm{xy} \mathbf{\nabla} \times \mathbf{a}$.
From Eq.~\eqref{anomalycancelation}, one can deduce:
\begin{subequations}
\begin{alignat}{4} 
 \hspace{-0.28cm} \mathcal{L}_\mathrm{eff}^{\partial \Omega}   \! & = \mathcal{L}_\mathrm{eff}^\mathrm{L} \,  \delta \left( y-y_{_\mathrm{L}} \right) + \mathcal{L}_\mathrm{eff}^\mathrm{R} \, \delta \left( y-y_{_\mathrm{R}} \right), \nonumber  \\
   \hspace{-0.28cm}  \mathcal{L}_\mathrm{eff}^{\rfrac{\mathrm{L}}{\mathrm{R}} } \!  &=   \chi^\dagger \, \im \left( \partial_t \mp  \frac{\mathrm{h}}{\mathrm{e}^2} \kappa_{_\mathrm{QAH}} \, D_\mathrm{x} \right) \chi \label{edgeliquidQAH} \\
    &  + \, \xi_{0}^\dagger  \, \im \left( \partial_t \mp \frac{\mathrm{h}}{\mathrm{e}^2} \kappa^{n=0}_{_\mathrm{QH}} \ D_\mathrm{x} \right) \xi_{0} \nonumber \\
    &\times  \Theta\left(\vert\mu+D/l_H^2 \vert-\vert M-B/l_H^2 \vert\right) \label{edgeliquidn0}  \\
& + \sum_{\substack{n =1 \\ \, s=\pm }}^{\infty} \!  s \, \xi_{n}^\dagger \,  \im \left( \partial_t \mp \frac{\mathrm{h}}{\mathrm{e}^2} \kappa^n_{_\mathrm{QH}} \, D_\mathrm{x} \right) \xi_{n}  \Theta \left[ s (\mu -  E_n^s ) \right], \label{edgeliquidn} 
\end{alignat}
\label{edgeliquids}
\end{subequations}

\vspace{-0.25cm}

\noindent
where $\chi$  $(\xi_n)$ defines QAH  (QH) edge states and $D_\mathrm{x}\!\equiv\!\partial_\mathrm{x}\!+\!\im \mathrm{e}\, a_\mu /\hbar$.  Equation~\eqref{edgeliquidQAH} is linked to Eq.~\eqref{kappa1} and characterizes QAH edge states, persisting in orbital fields. The QAH edge states are not bound to a specific LL (no Heaviside function) but instead bridge the gap between valence and conduction band. This finding is in accordance with our band structure calculations, shown in Fig.~\ref{fig:QAHEvolution}.  Since Eq.~\eqref{edgeliquidQAH} is connected to the spectral asymmetry $\eta_{_H}$, charge pumping via anomaly cancellation can occur from the QAH edge states into any LL. This pumping mechanism is therefore a signature of the parity anomaly and can, in general, exist until the Dirac mass gap is eventually closed at the critical field  [Eq.~\eqref{kappa1}]
\begin{align} \label{eq:gapClosing}
H_{\mathrm{crit}} = \sign (\mathrm{e} H)\,\dfrac{\hbar}{\mathrm{e}} \dfrac{M}{B} \ .
\end{align}
In contrast, \cref{edgeliquidn0,edgeliquidn} are related to \cref{kappa2,kappa3} and define QH edge states. 
These states are bound by  single LLs and charge flow appears only between edge states and their associated LL. 

\textit{Charge pumping.} To highlight the differences in the charge pumping between QAH and QH phases, we consider here an impurity-free system and comment on (in)elastic scattering effects in the next section.  
We simulate the evolution of the charge distribution as a function of the orbital field by solving the time-dependent Schr\"{o}\-din\-ger equation. As in typical experiments, we keep the total charge (not chemical potential) constant in our simulations \cite{Novik05,Bruene2011,Baum2014,Budewitz17}. In particular, we consider a vector potential  $\mathbf{A}(t)=\mathbf{A}(t_\mathrm{i})+\mathbf{a}(t)$ with $t\!\in\![t_\mathrm{i}\!=\!0 ,t_\mathrm{f}]$, where $\mathbf{A}(t_\mathrm{i})$ is a time-independent background field and $\mathbf{a}(t)=-yH(t)\mathbf{e}_\mathrm{x}$ is a time-dependent perturbation. At initial time $t_\mathrm{i}$, the system is described by the solutions of the Schr\"{o}dinger equation  $\vert \psi_{j,k_\mathrm{x}}(t_\mathrm{i})\rangle$, where $j$ labels the $j$-th subband. For $t>t_\mathrm{i}$, the perturbation is switched on and each initially occupied state, with $j \leq j_\mathrm{max}$ and $k \leq k_\mathrm{max}$, evolves under unitary time evolution to $\vert \psi_{j,k_\mathrm{x}}(t)\rangle$ \cite{Messia65}.  The quantities $j_\mathrm{max}$ and $k_\mathrm{max}$ are determined by the initial chemical potential $\mu$. 
Linearly increasing the orbital field with time,  we trace the occupation of states in each instantaneous spectrum, defined by the time-independent Schr\"{o}dinger equation $\mathcal{H}(t)\vert\phi_{i,k_\mathrm{x}}(t)\rangle=E_{i,k_\mathrm{x}}(t)\vert\phi_{i,k_\mathrm{x}}(t)\rangle$. 
Their occupation probabilities are determined by ${P_{i,k_\mathrm{x}}(t)= \sum_{j=0}^{j_\mathrm{max}}\!\,\vert\langle\psi_{j,k_\mathrm{x}}(t)\vert\phi_{i,k_\mathrm{x}}(t)\rangle\vert^2}$ [cf. App.~D].
\begin{figure}[!t]
	\centering
	\includegraphics[width=0.96\columnwidth]{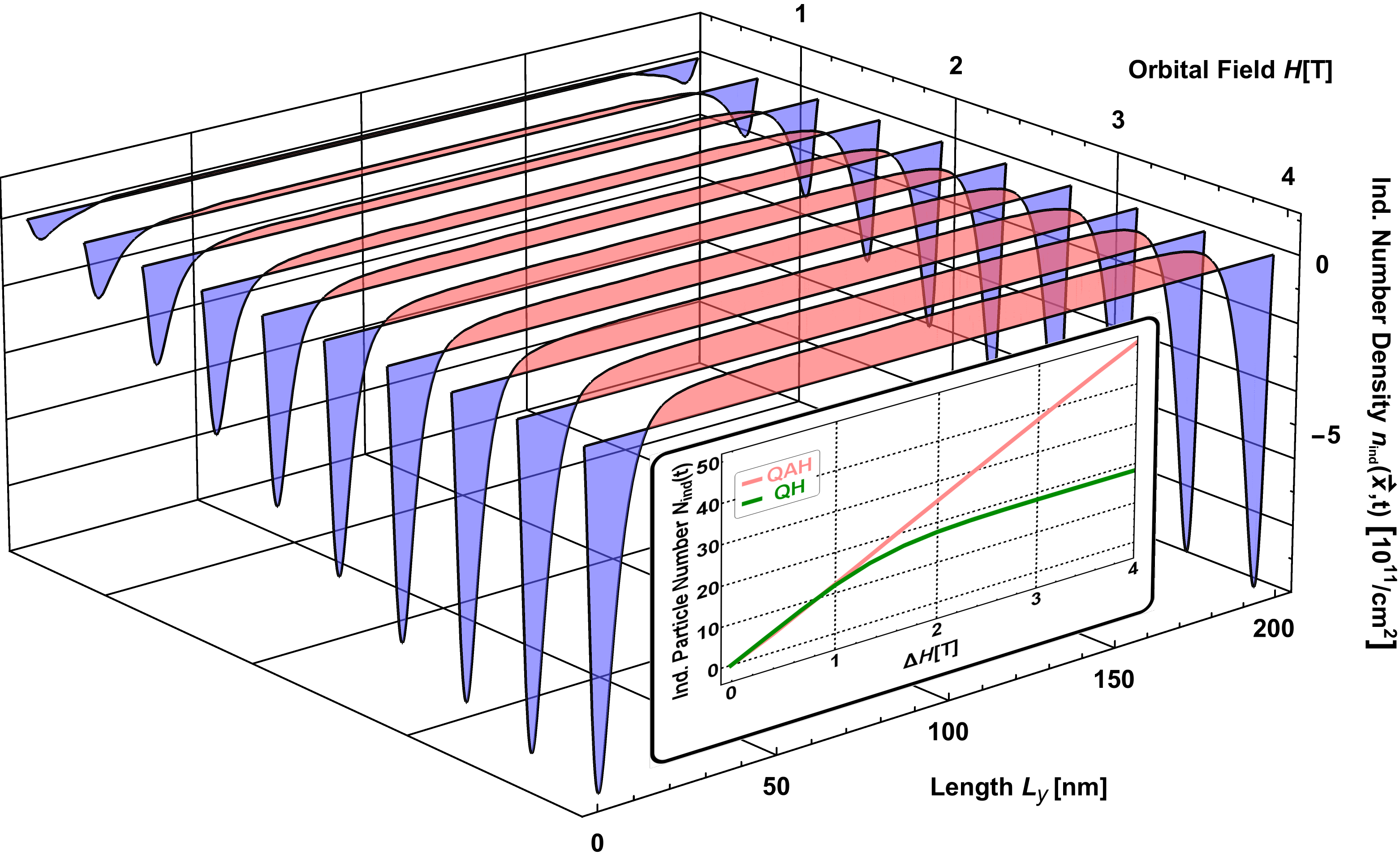}
	\caption{\label{fig:QAHEvolutionCharge} Evolution of $n_\mathrm{ind}(\mathbf{x},t)$ in orbital fields, corresponding to Figs.~\ref{fig:QAHEvolution}(a)-(c). 
An increase of $H$ causes charge depletion (blue) at the edges and charge accumulation (red) in the bulk. The inset compares the induced bulk particle number $N_\mathrm{ind}(t) = \int \! \mathrm{d}\mathbf{x} \ n_\mathrm{ind}(\mathbf{x},t)$  between the QAH (red)  and the QH phase (green).
}
\vspace{-0.2cm}
\end{figure}
At $t_\mathrm{i}$, the ground state for $(I)$ the QAH phase is determined by  $\mathbf{A}(t_\mathrm{i})=0$ with $\mu$ located at the Dirac point [Fig.~\ref{fig:QAHEvolution}(a)], whereas for $(II)$  the QH phase a finite background field $\mathbf{A}(t_\mathrm{i})=-yH_\mathrm{0}\mathbf{e}_\mathrm{x}$ has to be applied  and  $\mu$ is placed  above the first LL [Fig.~\ref{fig:QAHEvolution}(d)]. 
The numerical results, presented in Figs.~\ref{fig:QAHEvolution} and  \ref{fig:QAHEvolutionCharge}, are independent of the time scale in which $H(t)$ is ramped up, provided that $t_\mathrm{f}^\mathrm{min} \! \ll \!  t_\mathrm{f} \!  \ll \!  t_\mathrm{f}^\mathrm{max}$. The lower bound prevents excitations across bulk gaps $E_\mathrm{g}$ and is therefore determined by $t_\mathrm{f}^\mathrm{min} \equiv \hbar / E_\mathrm{g} \sim 10^{-13} \, \mathrm{s}$. 
For $H>H_\mathrm{scat}$, the upper bound comes from the requirement to overcome hybridization gaps forming between the QAH edge states and bulk LLs.
As long as these hybridization gaps are finite size gaps, exponentially suppressed by the system size, $t_\mathrm{f}^\mathrm{max}$ tends to infinity [App.~D].

Let us now discuss the numerical results, starting with the QAH phase under initial condition $(I)$. 
Increasing $H(t)$ with time,  the occupation of the eigenstates and the induced charge carrier density ${j^0_{\mathrm{ind}}(\mathbf{x},t)=-\mathrm{e}\,n_\mathrm{ind}(\mathbf{x},t)}$ evolve as shown in Figs.~\ref{fig:QAHEvolution}(a)-(c) and Fig.~\ref{fig:QAHEvolutionCharge} with $n_\mathrm{ind}(\mathbf{x},t)=\sum_{i,k_\mathrm{x}} P_{i,k_\mathrm{x}}(t)|\phi_{i,k_\mathrm{x}}(\mathbf{x},t)|^2-n_\mathrm{back}$, where $n_\mathrm{back}$ ensures that $n_\mathrm{ind}(\mathbf{x},t_i)\!=\!0$.
Starting from a flat (zero) charge density distribution, an increase of $H(t)$ causes a net charge flow from the QAH edge states (charge depletion) into all valence band LLs (charge accumulation). Since our system is a bulk insulator, this redistribution of charges is driven by polarization effects.
As a function of the orbital field 
all occupied wave functions shift their spectral weight, effectively giving rise to the charge redistribution shown in Fig.~\ref{fig:QAHEvolutionCharge}.
During this process, all valence band LLs, including the $n=0$ LL, remain filled.  As illustrated in the inset of Fig.~\ref{fig:QAHEvolutionCharge}, this causes a linear increase of the bulk charge with $j^0_\mathrm{ind}=\sigma_\mathrm{xy}\nabla \times \mathbf{a}=\kappa_{_\mathrm{QAH}} H(t)$. Since this type of pumping is  bound to the existence of the QAH edge states, it can only exist for  ${H<H_\mathrm{crit}}$ [Eq.~\eqref{eq:gapClosing}].	
These results are consistent with our conclusions based on the Callan-Harvey mechanism following from Eq.~\eqref{anomalycancelation}.

In contrast, our results  for the QH phase under initial condition $(II)$ are shown in  Fig.~\ref{fig:QAHEvolution}(d) and in the inset of Fig.~\ref{fig:QAHEvolutionCharge}. In agreement with our field-theoretical approach, we find that the bulk charge originates purely from the associated QH edge states, implying a saturation of the charge accumulation already for small orbital fields. This is therefore further evidence that the QAH edge states are related to a distinct CS term, which is connected to the spectral asymmetry $\eta_{_H}$ and not to a single LL.

\begin{figure}[!t]
	\centering
	\includegraphics[width=1\columnwidth]{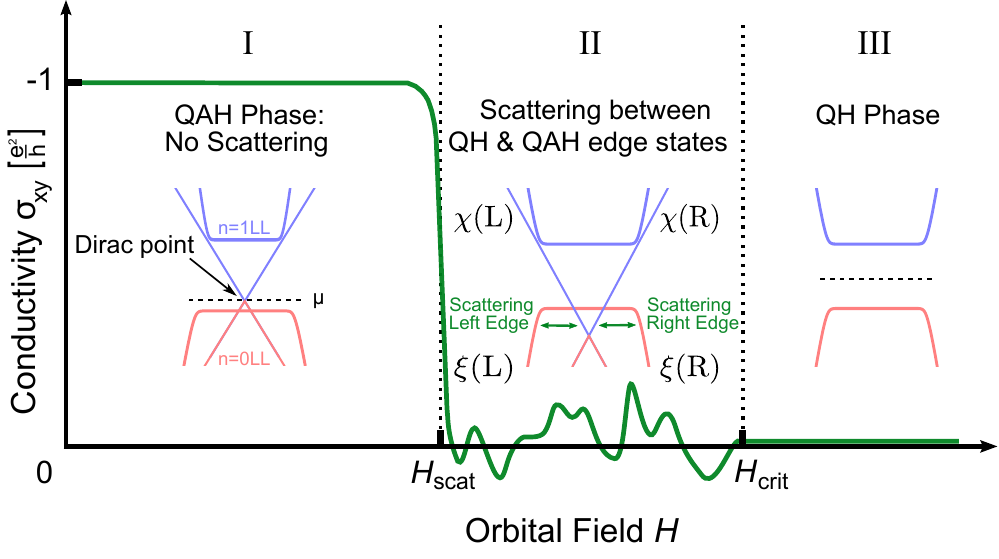}
	\caption{\label{fig:crossover} Schematic evolution of $\sigma_\mathrm{xy}$ for a QAH insulator in orbital fields in the presence of disorder. Insets schematically illustrate the underlying band structure according to Fig.~1(a)-(c) (same color code). In region \RM{2}, scattering processes between counterpropagating QH $\xi(L/R)$ (red) and QAH $\chi(L/R)$ (blue) edge states allow for momentum and energy relaxation.}
\vspace{-0.3cm}
\end{figure}

\textit{Experimental signatures.} 
We have so far considered an impurity-free system.
What are consequences of taking disorder and, therefore, (in)elastic scattering into account?
As long as the Dirac point is above the $n\!=\!0$ LL, i.e. for $H<H_\mathrm{scat}$, the system is in its ground state.
Scattering cannot cause relaxation of the induced bulk charge and, hence, disorder cannot affect the results of Figs.~\ref{fig:QAHEvolution}(b) and \ref{fig:QAHEvolutionCharge}. The hallmark of the QAH effect is a quantized Hall plateau with $\sigma_\mathrm{xy}=\kappa_{_\mathrm{QAH}}$ whose length scales with $H_\mathrm{scat} \sim L_\mathrm{y}^{-1}$. This is depicted by region \RM{1} in Fig.~\ref{fig:crossover} and follows from $g_\mathrm{eff}\sim L_\mathrm{y}$ [App.~A].   For $H>H_\mathrm{scat}$, the system is driven into a state with no common chemical potential, whose signature is a selective population of states (charge inversion), shown in Fig.~\ref{fig:QAHEvolution}(c). This charge inversion is protected by momentum conservation, since direct relaxation processes, such as spontaneous emission, are exponentially suppressed by the spatial localization of the wave functions. 
However, since realistic systems are rather imperfect, in(elastic) scattering between occupied QH and unoccupied QAH edge states facilitate  momentum and energy relaxation as indicated by region \RM{2} in Fig.~\ref{fig:crossover}. As a result, the charge inversion relaxes eventually, until a common chemical potential has set in. In this new ground state, counterpropagating QAH and QH edge states coexist at a single boundary. For instance in the inset of region \RM{2}, at the right boundary, the QAH edge state has a positive velocity, 
while the QH edge state has a negative velocity. 
Similarly to Ref.~\cite{Wang13}, which uses the Landauer-B\"{u}ttiker formalism \cite{Buettiker86,Buettiker88}, we expect deviations from a perfectly quantized Hall plateau arising from  scattering between QH and QAH edge states. When the transmission probability $T_{i,j}$ between contacts $i$ and $j$ on a typical Hall bar is symmetric, meaning $T_{i,i+1} = T_{i+1,i}$, we expect a $\sigma_\mathrm{xy}=0$ plateau [App.~E]. If scattering processes between the coexisting edge states microscopically differ on both edges of the Hall bar,  deviations from a perfect quantization arise (wiggly line in Fig.~\ref{fig:crossover}). In contrast for $T_{i,i+1} \neq T_{i+1,i}$, the average value of $\sigma_\mathrm{xy}$ can significantly deviate from zero. Such direction-dependent transmission probabilities can result from a large charge puddle density [App.~E] (diffusive regime) which is typically present in large (Hg,Mn)Te Hall bars \cite{Vayrynen13,Lunczer19,Roth09}. Finally for $H>H_\mathrm{crit}$, the Dirac mass gap is closed and $\sigma_\mathrm{xy}$ vanishes as indicated by region \RM{3} in Fig.~\ref{fig:crossover}.

\textit{Realization.} Typical materials in which this crossover should be observed include (Hg,Mn)Te/CdTe quantum wells,  described by the BHZ model \cite{Bernevig06,Liu08,Rothe10,Budewitz17}. In the discussion above, we assumed that the spin-down block of the BHZ model is trivial and, hence, does not qualitatively affect the discussed physics. Nevertheless, analogous equations for the spin-down block can be derived replacing $\left(M,B\right)\! \rightarrow \! \left(-M,-B\right)$. Zeeman ($g_\mathrm{z}$) and exchange ($G_\mathrm{ex}$) terms  can be incorporated, replacing $M \rightarrow M \pm g(H)$, where $g(H) \equiv g_\mathrm{z} H + G_\mathrm{ex}(H)$ \cite{Liu08} and $+(-)$ applies to the spin-up(down) block.
In the full BHZ model, $g\!\neq\!0$ breaks time-reversal symmetry and drives the system into the QAH phase if ${(M \! + \! g \! - \! B/l_H^2)(M \!-\!g\!-\!B/l_H^2)\!<\!0}$, extending the definition of QAH insulators to orbital fields \cite{Liu08}. Since the exchange interaction in (Hg,Mn)Te is paramagnetic \cite{Furdyna88},  a finite orbital field is needed to drive the system into the QAH phase. In the full BHZ model, the spin-down block causes an additional transition from the QSH phase to region \RM{1} in Fig~\ref{fig:crossover}. 
In Bi-based QAH insulators, one should be able to  observe similar transitions as shown  in Fig.~\ref{fig:crossover}, given that signatures of both the QH and  the QAH effect are observed at relatively small orbital fields \cite{Chang13,Checkelsky14,Moodera15,Bestwick15}.

\textit{Conclusions.} 
The field theoretical analysis of QAH insulators in orbital fields allows us to explain the very unconventional findings in band structure calculations based on the parity anomaly. In particular, we reveal three novel transport features which are all fundamentally based on the parity anomaly:
A violation of the Onsager relation, a peculiar type of charge pumping with increasing orbital field, and, for large fields, the emergence of counterpropagating QH and QAH edge states. Together these signatures highlight the different physical origin of topology of QH and QAH phases, making them distinguishable even though they are described by the same Chern number.  As a fingerprint of these features, we predict a transition from $\sigma_\mathrm{xy}= -\mathrm{e}^2/\mathrm{h}$ (QAH effect) to a noisy QH plateau with increasing orbital fields, whose average value depends on details of the QH-QAH edge state scattering. The experimental verification of our theoretical predictions in (Hg,Mn)Te quantum wells is underway \cite{Budewitz17}. In future, it would be interesting to study signatures of quantum anomalies beyond the BHZ model and analyze  microscopic signatures of  counterpropagating QH and QAH edge states therein.

\begin{acknowledgments}
\textit{Acknowledgments.} 
We thank B.~Scharf, R.~Meyer, B.~Trauzettel, B.~A.~Bernevig, F.~Wilczek,  C.~Br\"{u}ne, J.~S.~Hofmann, J.~Erdmenger, C.~Morais Smith, and W.~Beugeling
 for useful discussions. 
We acknowledge financial support through the Deutsche Forschungsgemeinschaft (DFG, German Research Foundation), project-id 258499086 - SFB 1170 'ToCoTronics', the Free State of Bavaria (Elitenetzwerk Bayern IDK "Topologische Isolatoren" and the Institute for Topological Insulators), and through the W\"urzburg-Dresden Cluster of Excellence on Complexity and Topology in Quantum Matter - ct.qmat \mbox{(EXC 2147, project-id 39085490)}.
\end{acknowledgments}

\bibliographystyle{apsrev4-1}

%

\appendix
\section{Appendix A: Remnant of QAH effect in orbital fields} \label{AppendixA}
In this section, we are going to show that the band inversion of a non-trivial Chern insulator survives up to orbital fields of
\begin{align} \label{eqA:criticalField}
H_{\mathrm{crit}}=\sign (\mathrm{e} H)\,\frac{M\, \phi_\mathrm{0}}{B\, 2 \pi },
\end{align}
where $\phi_\mathrm{0}=\mathrm{h}/\mathrm{e}$. In Section A.1, this relation is first proven from a pure bulk approach, while  in Section A.2, we supply further evidence using an edge approach. A more rigorous  argument will be presented in App.~B.
\subsection*{Bulk Approach \label{sec:A1}}
In this appendix, we  take a closer look on the features of an inverted band structure surviving in orbital fields. In particular, we study a 2D Chern insulator, which is denoted by \cite{Bernevig06}
\begin{align}
\! \! \mathcal{H}=\left(M-B k^2\right)\sigma_\mathrm{z}-D k^2 \sigma_{0}+A\left(k_\mathrm{x} \sigma_\mathrm{x}-k_\mathrm{y} \sigma_\mathrm{y} \right) \ ,
	\label{eqA:ChernHamiltonAp}
\end{align}
where all parameters have been discussed in the main text. For simplicity, let us adopt this Hamiltonian in the pseudospin basis of the spin-up block of $\mathrm{Hg}_{1-y}\mathrm{Mn}_y\mathrm{Te}$, i.e., $\left\{|E1,\uparrow\rangle,|H1,\uparrow\rangle\right\}$. The energy spectrum of Eq.~\eqref{eqA:ChernHamiltonAp} is determined by solving the corresponding Schr\"{o}dinger equation:
\begin{align}\label{eqA:bulkSpectrum}
E^\pm\left(k_\mathrm{x},k_\mathrm{y}\right)=-Dk^2\pm\sqrt{A^2k^2+\left(M-Bk^2\right)^2} \, .
\end{align}
\begin{figure}[!b]
	\centering
	\includegraphics[width=0.49\columnwidth]{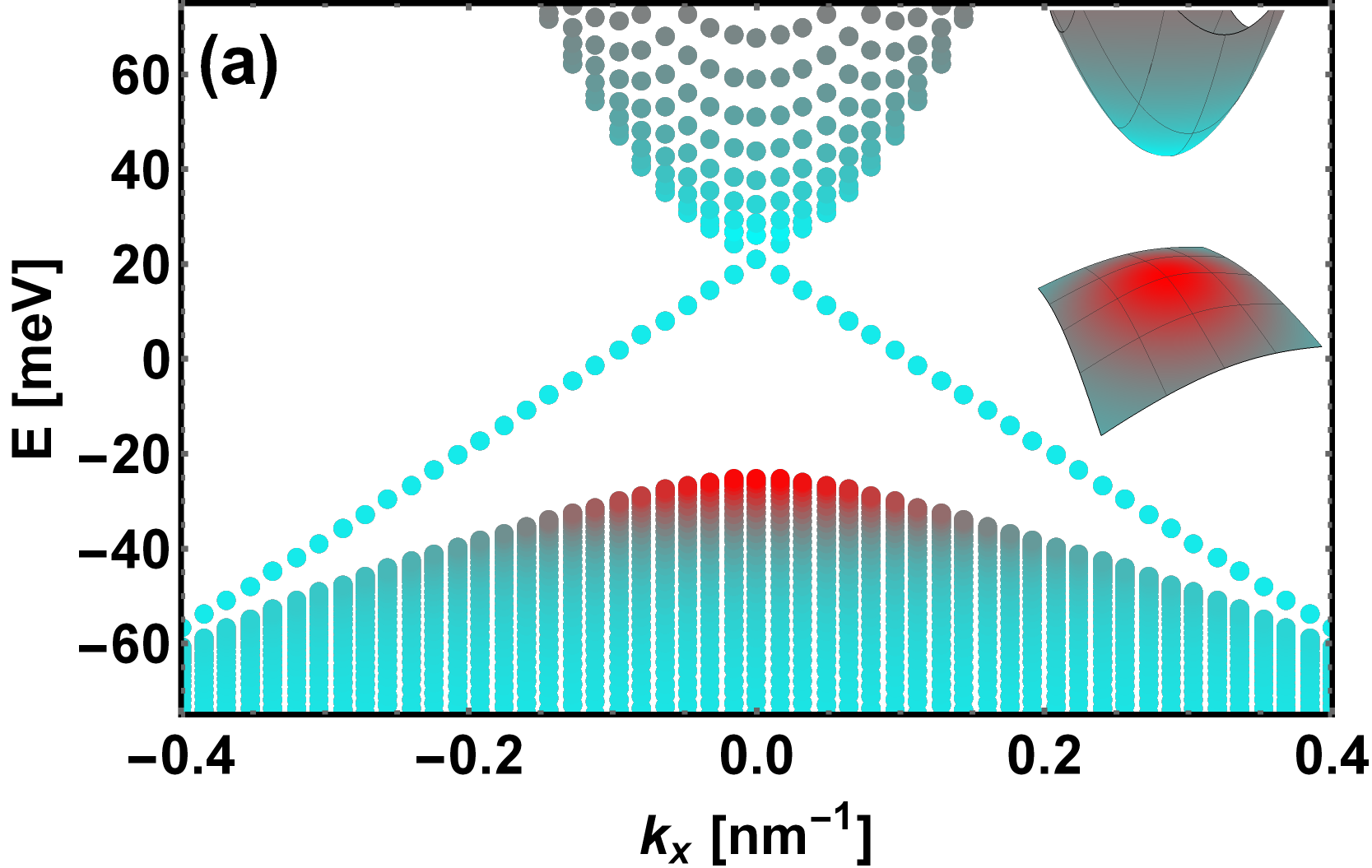}
	\hfill
	\includegraphics[width=0.49\columnwidth]{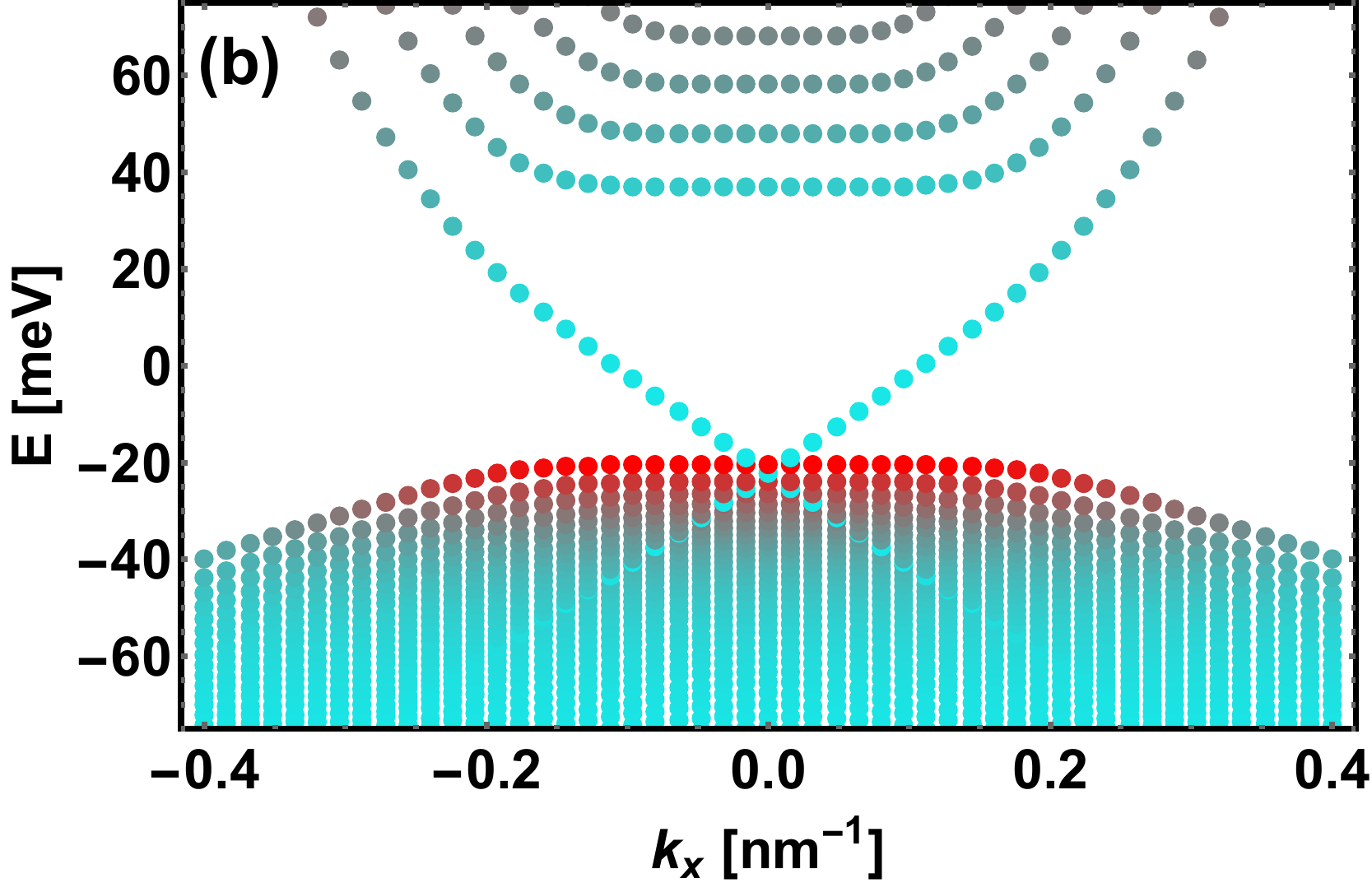}
	\caption{\label{figAp:InvertedBandStructure} Band structure of a Chern insulator on a stripe geometry with parameters chosen as in Fig. 1 of main text. The color code  indicates that the respective wave functions consist of a mixture of $E1$ (red) and $H1$ (cyan) components. In (a) the band structure is depicted at $H=0\;\mathrm{T}$. One clearly observes the band inversion close to the $\Gamma$-point. The inset shows the continuous bulk band structure as determined by Eq.~\eqref{eqA:bulkSpectrum}. In (b) the spectrum is shown for $H=1.5 \,\mathrm{T}$. }
\end{figure}
\noindent
The system has a non-trivial topology if $M/B>0$. In the following, we focus without loss of generality on $M,B<0$ \cite{Bernevig06}. A consequence  of this condition is that the band structure is inverted, meaning that the band edge energy of the $E1$-band lies below the band edge energy of the $H1$-band. An exemplary case is depicted in Fig.~\ref{figAp:InvertedBandStructure}(a), where the color code displays the pseudospin character of the wave functions highlighting the inverted band structure. In particular, Fig.~\ref{figAp:InvertedBandStructure}(a) depicts the band structure for a stripe geometry and only the inset shows the continuous bulk band structure as determined by Eq.~\eqref{eqA:bulkSpectrum}. In momentum space, we observe that the band structure is inverted close to the $\Gamma-$point, characterized by $M-Bk^2<0$, while it is  normally ordered for momenta satisfying $M-Bk^2>0$. This property results from a momentum dependent renormalization of the  Dirac mass $M$ by the effective mass $Bk^2$.  The band ordering is solely determined by the diagonal terms of the Hamiltonian. However, closing of the Dirac mass gap at finite momentum cannot be observed in band structure plots, since the off-diagonal terms cause hybridization between bulk bands. Nevertheless, the bulk band inversion is removed at  $k_\mathrm{crit}^2=M/B$, where $M-Bk^2=0$ implying that the renormalized Dirac mass changes its sign. 

\begin{figure}[!t]
\vspace{-0.5cm}
\begin{minipage}{1.0\columnwidth}
\includegraphics[width=.8\columnwidth]{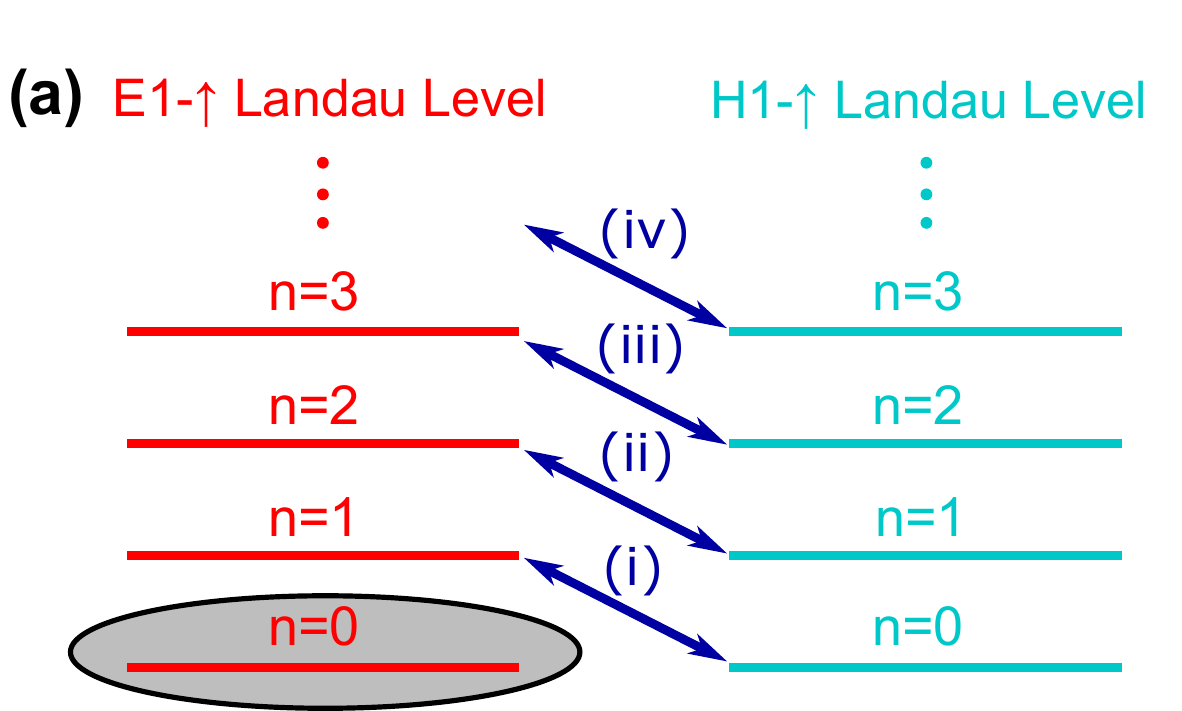} \vspace{0.5cm}
\includegraphics[width=0.95\columnwidth]{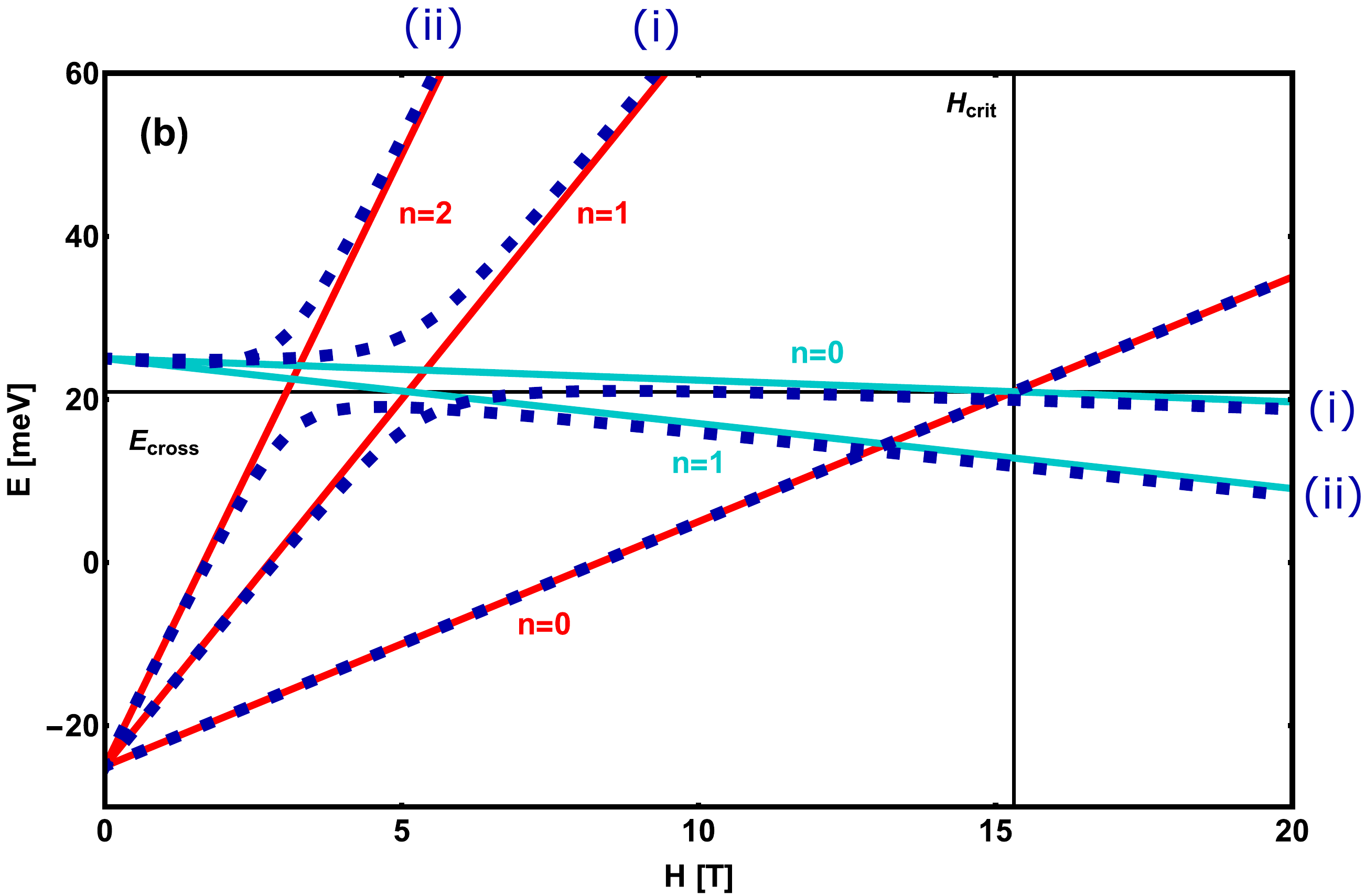}
\end{minipage}
\caption{\label{figAp:originLandauFan}Hybridization and gap closing of the bulk LL spectrum in orbital fields. Except for $A$, all parameters are taken from the caption of Fig.~\ref{figAp:InvertedBandStructure}. The hybridization between $E1$ and $H1$ bands in orbital fields is schematically demonstrated in (a), where red (cyan) indicates $E1$ ($H1$) - pseudospin character. In (b), the LL energies without coupling ($A=0$) are depicted by solid lines and are explicitly numerated by LL indices. Blue dotted lines show the evolution of LLs with hybridization for $A=45\;\mathrm{meVnm}$. }
\end{figure}

Now, we raise the question, what happens to the band inversion in the presence of an orbital field? As it is well known, an orbital field induces a parabolic confinement and increases (decreases) the energy of $E1$ ($H1$) - states independently on the orbital field direction. It should therefore counteract the band inversion. Since the band inversion is protected by the Dirac mass gap, a finite critical orbital field is needed to remove the inversion completely. In Fig~\ref{figAp:InvertedBandStructure}(b), the band structure is shown for $H=1.5$ T and, on the first glance, the color code seems to reveal that the band inversion is still present. In the following, we give a proof for Eq.~\eqref{eqA:criticalField} based on a simple analysis of the bulk LL spectrum.
Let us start by replacing the gauge-independent momentum operators, $\pi_i=k_i+ \mathrm{e}A_i/\hbar $, by ladder operators
\begin{align*}
&\pi_+\rightarrow\frac{\sqrt{2}}{l_H}\begin{cases} \ a^\dagger \text{ for } s>0 \\ -a \text{ for } s < 0 \end{cases}  \\
&\pi_-\rightarrow\frac{\sqrt{2}}{l_H}\begin{cases} \ \ \ \, a  \text{ for } s>0 \\ -a^\dagger  \text{ for } s < 0 \end{cases} \, ,
\end{align*} 
where $s=\text{sgn}(\mathrm{e}H)$ and $l_H=\sqrt{\hbar/\vert \mathrm{e} H \vert}$. An appropriate ansatz to solve the Schr\"{o}dinger equation is given by
\begin{align} \label{eqAp:landauLevelSpinorWFn}
	\psi_{n,k_\mathrm{x}}^\pm (y)& \propto
	\begin{cases}
			\begin{pmatrix}
		\left(M-\beta n-\frac{s \delta}{2} \pm \epsilon_n\right)\langle y \vert n,k_\mathrm{x}\rangle \\
		s\alpha \sqrt{n} \langle y \vert n-1,k_\mathrm{x}\rangle
		\end{pmatrix} & \mkern-13mu s>0\\
		\\
		\begin{pmatrix}
		\left(M-\beta n-\frac{s \delta}{2} \pm \epsilon_n \right)\langle y \vert n-1,k_\mathrm{x}\rangle \\
		s\alpha \sqrt{n} \langle y \vert n,k_\mathrm{x}\rangle
		\end{pmatrix} & \mkern-13mu s<0
	\end{cases}
\end{align}
and
\begin{align} \label{eqAp:landauLevelSpinorWFnZero}
	\psi_{0,k_\mathrm{x}}(y)& \propto
	\begin{cases}
		\begin{pmatrix}
		\langle y \vert 0, k_\mathrm{x} \rangle \\
		0
		\end{pmatrix} & s>0 \\
		\\
		\begin{pmatrix}
		0\\
		\langle y \vert 0, k_\mathrm{x} \rangle
		\end{pmatrix} & s<0 \ ,
	\end{cases}
\end{align}
where we neglected normalization constants for simplicity, $\alpha=\sqrt{2}A/l_H$, $\beta=2B/l_H^2$, and $\delta=2D/l_H^2$. The corresponding LL energies are  \cite{Konig08}
\begin{align}
	E_{n\neq0}^{\pm}&=-s\beta/2-n\delta \pm\epsilon_n,	\label{eqAp:nLLenergies} \\
	E_\mathrm{0}&=s\left( M-\beta/2 \right)-\delta/2, 	\label{eqAp:0LLenergies} 
\end{align}
where $\epsilon_n= \sqrt{ \alpha^2 n +(M-n\beta-s\delta/2 )^2 }$. A hallmark of a Dirac-like Hamiltonian is the relativistic structure of the LL spinors, following from the off-diagonal structure of the Hamiltonian. The $n=0$ LL is completely decoupled and, therefore, pseudospin polarized, while all other LLs (for $n>0$) are formed from hybridization between the $n$-th $E1$ and $(n-1)$-th $H1$ LLs \cite{Konig08}. This asymmetric coupling in orbital fields causes an asymmetry in the LL spectrum further discussed in App.~B and in Ref.~\cite{Niemi83,Schakel91}. A sketch of this hybridization process is shown in Fig.~\ref{figAp:originLandauFan}(a). Notice that without hybridization, the system would include two decoupled $n=0$ LLs.

Above, we have shown for $H=0$, that the diagonal elements in Eq.~\eqref{eqA:ChernHamiltonAp} characterize the band ordering in momentum space. These elements will also define the band ordering of the Chern insulator in orbital fields. To understand this in detail, let us first study the evolution of the $E1$ and $H1$ LLs without hybridization. In such a case, $A$ is effectively zero and the system is driven by its diagonal elements. This is illustrated by the solid lines in Fig.~\ref{figAp:originLandauFan}(b), where it is found that each pair of LLs with opposite pseudospin, but with the same LL index $n$, crosses at the same energy $E_\mathrm{cross}=-MD/B$. Such a LL crossing is a signature of an inverted band structure. The higher the LL index, the earlier a pair of states crosses each other in orbital fields. Without hybridization, the latest crossing would occur between the two ${n=0}$ LLs, exactly at the critical field given by Eq.~\eqref{eqA:criticalField}, i.e., at $H_\mathrm{crit}=\sign(\mathrm{e}H)M\phi_\mathrm{0} / (B 2\pi)$. This marks the point at which all $E1$ LLs are above $H1$ LLs. Hence for $H>H_\mathrm{crit}$, the band structure becomes normally ordered.

However, for a finite coupling parameter $A$, this crossing is hidden due to the hybridization between the ${n\!=\!1-\mathrm{E1}}$ LL and the ${n\!=\!0-\mathrm{H1}}$ LL as depicted by the blue dashed line in Fig.~\ref{figAp:originLandauFan}(b) (we have chosen here a smaller $A$ than in Fig.~\ref{figAp:InvertedBandStructure} to make the hybridization between adjacent LLs clearer). Following the discussion at $H=0$, the band ordering in orbital fields should be solely determined by the diagonal terms of the Hamiltonian. The coupling parameter $A$ hides the critical point at which the band structure becomes normally ordered. We therefore deduce that $H_\mathrm{crit}$  defines the critical orbital field at which the band inversion is completely removed.

To avoid any misunderstanding, we want to emphasize that the argumentation above has been given for a single block of the BHZ model. The discussed crossing should not be confused with the one, occurring in the full BHZ model. In this model, one observes a crossing between  two pseudospin polarized $n=0$ LLs \cite{Konig08}, where each level belongs to a distinct spin-block.  

\subsection*{Edge approach}
\begin{figure}[!t]
\includegraphics[width=0.98\columnwidth]{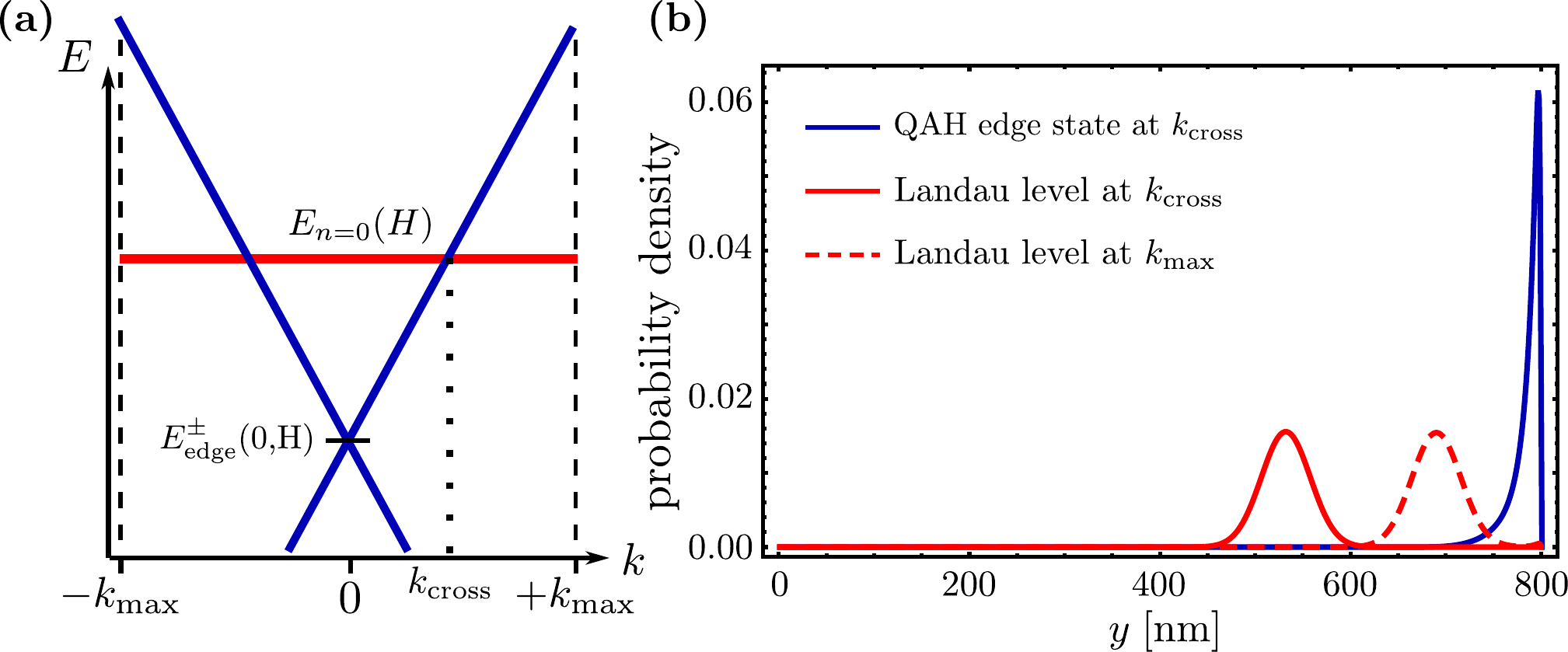}
\caption{\label{fig:qahEdgeLLHyb} Energy gaps between a QAH edge state (blue)  and the $n=0$ LL (red). (a) The sketch shows that the QAH edge states was shifted down in energy by the orbital field and crosses the flat bulk LL at $k_\mathrm{cross}$. (b) If $k_\mathrm{cross} \ll k_\mathrm{max}$, the wave functions of the bulk LL (solid red line) and of the QAH edge state (solid blue line) have an exponentially small overlap. Only if the crossing of the QAH edge state and the bulk LL happens close to $k_\mathrm{max}$ (wave function indicated by red dashed line), they would start to hybridize causing a finite gap in the spectrum.}
\end{figure}

The upper proof is based on a pure bulk calculation. Here, we show that the result for $H_\mathrm{crit}$ can be reproduced focusing on the evolution of the QAH edge states in orbital fields.  As described in the main text, the QAH edge states are successively lowered in energy as we increase the orbital field (for $\text{sgn}(\mathrm{e} H)>0$). For simplicity, Fig.~\ref{fig:qahEdgeLLHyb} sketches this scenario by only taking into account the flat bulk $n=0$ LL, as well as the QAH edge states.  With this simplified model, we are aiming to determine an upper limit until which the QAH edge states can survive in orbital fields. By survival of the QAH edge states, we mean that even for $H \neq 0$ the QAH edge states and bulk LLs remain decoupled up to finite size gaps, exponentially vanishing in the limit $L_\mathrm{y} \rightarrow \infty$.

First, let us take a closer look on the properties of the $n=0$ LL, whose energy is given by  Eq.~\eqref{eqAp:0LLenergies}. The degeneracy of this level and therefore its width in momentum space increases linearly with $H$:
\begin{align} 
 k_\mathrm{max}= \dfrac{\mathrm{e} \, H \, L_\mathrm{y}}{2\hbar}, \nonumber
\end{align}
where $2 \, k_\mathrm{max}$ is the full width of the LL [Fig.~\ref{fig:qahEdgeLLHyb}]. 
The associated wave functions $\psi_{0,k_\mathrm{x}}(y)$ of the zeroth LL are each centered at $y(k_\mathrm{x})=l_H^2 k_\mathrm{x}$ and their spatial width decreases linearly as we increase the orbital field. 

Now, let us analyze the evolution of the QAH edge states in orbital fields for which an analytic expression   was derived by Zhou et al. \cite{Zhou08}:
\begin{align}
E_\mathrm{edge}^\pm\left(k_\mathrm{x},H\right)=E_\mathrm{D}(0) - \mu_\mathrm{B} g_{\mathrm{eff}}\left(L_\mathrm{y} \right) H \pm \hbar v_\mathrm{x} k_\mathrm{x} \ , \label{eqA:qahEdgeState}
\end{align}
where $v_\mathrm{x}$ is the edge state velocity, $\mu_\mathrm{B}$ is the Bohr magneton and  $E_\mathrm{D}(0)$ is the energy of the Dirac point at $H=0$. The effective g-factor  reads
\begin{align*} 
g_\mathrm{eff}(L_\mathrm{y})= \mathrm{m}_0 \, v_\mathrm{x} \hbar^{-1} \left[ L_\mathrm{y}- \lambda_1^{-1}- \lambda_2^{-1}-2 (\lambda_1 + \lambda_2)^{-1}  \right],
\end{align*}
where $\lambda_{1,2}$ are the decay length scales of the edge state and $\mathrm{m}_0$ is the bare electron mass. For $L_\mathrm{y}\gg\lambda_{1,2}^{-1}$, we  further simplify  $g_\mathrm{eff}(L_\mathrm{y}) \approx \mathrm{m}_0 v_\mathrm{x} \hbar^{-1} L_\mathrm{y}$. 

The crossing between the zeroth LL and the QAH edge states in momentum space, as shown in Fig.~\ref{fig:qahEdgeLLHyb} (a), is denoted by $k_\mathrm{cross}$. It moves to larger momentum values for increasing orbital fields. This is due to the fact that the QAH edge states are pushed down in energy while the $n=0$ LL is pushed up in energy as we increase $H$. As schematically shown in Fig.~\ref{fig:qahEdgeLLHyb}(b) (solid lines), the wave functions of the QAH edge states and the bulk LL are protected from hybridization due to their strong spatial localization. For $k_\mathrm{cross} \ll k_\mathrm{max}$, their overlap is exponentially small such that their hybridization gap is a finite size gap. Only if the crossing occurs close to the maximal width of the LL, bulk wave functions start to strongly overlap with the edge states and energy gaps, larger than finite size gaps, emerge. The maximal momentum above which the QAH edge states start to hybridize with bulk LLs is therefore determined by $k_\mathrm{cross}=k_\mathrm{max}$. Solving the equation $E_\mathrm{edge}(k_\mathrm{max},H_\mathrm{crit})=E_{n=0}(H_\mathrm{crit})$ for $H_\mathrm{crit}$, we find that the critical orbital field, above which the QAH edge states must start to hybridize strongly with bulk LLs, is again given by Eq.~\eqref{eqA:criticalField}. In general, the QAH edge states can therefore survive as long as remnants of the band inversion in orbital fields survive. Nevertheless note that hybridization gaps, larger than finite size gaps,  might occur already for $H<H_\mathrm{crit}$ as the given proof displays only an upper limit. This means that, for instance, higher order terms in $k_\mathrm{x}$ and $H$ in Eq.~\eqref{eqA:qahEdgeState} can become important. 

Furthermore, we use Eq.~\eqref{eqA:qahEdgeState} to compute the critical orbital field at which the Dirac point enters the valence band. This critical field was denoted by $H_\mathrm{scat}$ in the main text, since above this value scattering processes between counterpropagating QH and QAH edges states become possible. In order to derive an analytic expression for this field, we resolve  $E_\mathrm{edge}^{+}(0,H_\mathrm{scat})=E_{n=0}(H_\mathrm{scat})$ for $H_\mathrm{scat}$ which results in
\begin{align*}
H_\mathrm{scat}=\frac{M\left(B+D\right)}{B\left(B+D\right)\frac{\mathrm{e}}{\hbar}-\frac{\mathrm{e}}{2}Bv_\mathrm{x}\hbar L_\mathrm{y}}\propto\frac{1}{L_\mathrm{y}}.
\end{align*}

\section{Appendix B: Charge counting and effective field theory of Chern insulators}
In this appendix, we present a more detailed calculation on how to derive the bulk charge and the effective bulk Lagrangian [Eq.~(3) of main text] of a gapped Chern insulator in orbital fields. We take again Eq.~\eqref{eqA:ChernHamiltonAp} as our starting point and focus for simplicity on the particle-hole symmetric case,  $D=0$. In the literature, similar calculations were only performed for massive (2+1)D quantum electrodynamics  \cite{Niemi83,Niemi84,Niemi85,Boyanovsky86,Schakel91}, i.e., a Chern insulator without a $Bk^2$-term. 

As a Dirac-like system, a Chern insulator (continuum model) has an infinite Dirac sea, which causes infinities in many physical observables. To obtain finite results, those infinities have to be carefully subtracted (renormalized). For $H=0$, this can be achieved by the physical requirement that the fermion number needs to vanish if the chemical potential is located at the charge neutrality point $E_\mathrm{z}$. For a particle-hole symmetric Chern insulator, the charge neutrality point lies in the middle of the bulk gap at $E_\mathrm{z}=0$. Physically, we satisfy this constrain by choosing antisymmetrization as the appropriate operator ordering for the (normal ordered) bulk fermion number operator \cite{Niemi85}:
\begin{align}\label{eqA:fermionNoOpDef}
N = \frac{1}{2}\int_S \mathrm{d}\mathbf{x}\;\sum_{\alpha}\left[\psi^\dagger_\alpha(\mathbf{x}),\psi_\alpha(\mathbf{x})\right],
\end{align}
where $\psi(\mathbf{x})$ is a field operator (two component spinor). To calculate the bulk fermion number, the renormalization scheme, as given by Eq.~\eqref{eqA:fermionNoOpDef}, must be maintained for $H \neq 0$. In this case, the field operators can be expanded in terms of the normalized LL spinors of the conduction band $u_{n,k_\mathrm{x}}(\mathbf{x})= e^{i k_\mathrm{x} x}\psi_{n,k_\mathrm{x}}^+(y)$, as well as the valence band $v_{n,k_\mathrm{x}}(\mathbf{x})= e^{i k_\mathrm{x} x}\psi_{n,k_\mathrm{x}}^-(y)$, which have been already defined in Eqs.~\eqref{eqAp:landauLevelSpinorWFn} and \eqref{eqAp:landauLevelSpinorWFnZero}:
\begin{align}
	\psi \left(\mathbf{x}\right)=\sum_{k_\mathrm{x},n} b_{n,k_\mathrm{x}}u_{n,k_\mathrm{x}}(\mathbf{x})+\sum_{k_\mathrm{x},n} d_{n,k_\mathrm{x}}^\dagger v_{n,k_\mathrm{x}}(\mathbf{x})\label{eqA:fieldOp} \ .
\end{align}
Here $b_{n,k_\mathrm{x}}$ destroys an electron in the $n$-th conduction band LL with momentum $k_\mathrm{x}$, and $d^\dagger_{n,k_\mathrm{x}}$ creates a hole in the $n$-th valence band LL with momentum $k_\mathrm{x}$. The LL energies are determined by Eqs.~\eqref{eqAp:nLLenergies} and \eqref{eqAp:0LLenergies}. The zeroth LL is unique since it is either part of the valence or conduction band. As a result, for $E_{n=0}>E_\mathrm{z}$ the first sum in Eq.~(\ref{eqA:fieldOp}) runs from $n=0\ldots\infty$ and the second sum from $n=1\ldots\infty$. The situation is reversed, if the zeroth LL is located at an energy $E_{n=0}<E_\mathrm{z}$. 
All fermionic operators fulfill conventional anti-commutation relations
\begin{align}
	&\left\{b_{n,k_\mathrm{x}},b_{m,q_\mathrm{x}}^\dagger\right\}=\delta_{n,m}\delta_{k_\mathrm{x},q_\mathrm{x}} \, , \nonumber \\
	&\left\{d_{n,k_\mathrm{x}},d_{m,q_\mathrm{x}}^\dagger\right\}=\delta_{n,m}\delta_{k_\mathrm{x},q_\mathrm{x}} \, .
	\label{eqA:fermionicCom}
\end{align}
Inserting now Eq.~\eqref{eqA:fieldOp} into Eq.~\eqref{eqA:fermionNoOpDef} gives
\begin{align} \label{eqA:fullParticle}
N=&\frac{1}{2}\left(\sum_{k_\mathrm{x},n}[b_{n,k_\mathrm{x}}^\dagger,b_{n,k_\mathrm{x}}]+\sum_{k_\mathrm{x},n}[d_{n,k_\mathrm{x}},d_{n,k_\mathrm{x}}^\dagger]\right)\nonumber\\
=& N_0-\eta_{_H}/2 \ , 
\end{align}
where we made use of Eq.~(\ref{eqA:fermionicCom}). Here, $N_0$ and the, so called, spectral asymmetry $\eta_{_H}$ are given by \cite{Niemi85}
\begin{align}
	N_0 = &\sum_{k_\mathrm{x},n} b_{n,k_\mathrm{x}}^\dagger b_{n,k_\mathrm{x}}-\sum_{k_\mathrm{x},n}d_{n,k_\mathrm{x}}^\dagger d_{n,k_\mathrm{x}}, \label{eqA:bareCharge}\\
	\eta_{_H} = &\sum_{E>E_\mathrm{z}} 1-\sum_{E<E_\mathrm{z}} 1=\sum_{E}\text{sgn}\left(E-E_\mathrm{z}\right).
	\label{eqA:spectralAsym}
\end{align}
The spectral asymmetry $\eta_{_H}$ quantifies the asymmetry of the entire eigenvalue spectrum. This means that it counts the difference in the amount of states between valence and  conduction band. It is a topological quantity since it is invariant under small, local perturbations \cite{Niemi85}.

At $H=0$, the spectral asymmetry $\eta_{_H}$ vanishes due to the underlying particle-hole symmetry. This argumentation is however not valid in an orbital field where this symmetry is violated. Here every summand in Eq.~(\ref{eqA:spectralAsym}) contributes to $\eta_{_H}$, since there is no symmetry argument which allows us to cancel summands from the first with the second sum. Due to the fact that $\eta_{_H}$ consists of two infinite sums, which are separately divergent, a regularization scheme has to be introduced. Here, we make use of a heat-kernel regularization \cite{Nakahara2003},
\begin{align}\label{eqA:heatKernel}
	n\geq 0: \quad1\rightarrow e^{-\kappa \vert E_n\vert} \, ,
\end{align}
where $\kappa\!>\!0$ ensures the absolute convergence of both sums. At the end of the calculation, we will regain $\eta_{_H}$by taking the limit $\kappa\rightarrow 0^+$. Employing now Eq.~\eqref{eqA:heatKernel} to rewrite Eq.~\eqref{eqA:spectralAsym}, we obtain
\begin{align}
	\eta_{_H}(\kappa)=&\sum_{k_\mathrm{x},n=1} e^{-\kappa E_n^+}-\sum_{k_\mathrm{x},n=1} e^{\kappa E_n^-}+\sum_{k_\mathrm{x}} c e^{-\kappa \vert E_\mathrm{0}\vert}\nonumber\\
	=& \, n_\mathrm{0}\left( \sum_{n=1} e^{-\kappa E_n^+}-\sum_{n=1} e^{\kappa E_n^-}+c \, e^{-\kappa \vert E_\mathrm{0}\vert}\right) \, \label{eqA:stepEta}
\end{align}
where $c=\text{sgn}\left(\mathrm{e}H\right) \text{sgn}\left(M-\beta/2\right)$. Here, the last term marks the contribution of the zeroth LL and, in the second equality, we made use of the momentum independence of the eigenvalue spectrum to extract the LL degeneracy, given by $n_\mathrm{0}=\sum_{k_\mathrm{x}} 1 =S/(2\pi l_H^2)$. Here,   $S$ is the area of the system. To further simplify Eq.~\eqref{eqA:stepEta}, we  Taylor expand the eigenenergies for large $n$:
\begin{align*}
	E_n^\pm & =-s\frac{\beta}{2} \pm n |\beta| \sqrt{1+\frac{\alpha^2 n + M^2 - 2 M \beta n}{n^2 \beta^2} } \\
			& \approx -s \frac{\beta}{2} \pm \left[n |\beta|+\mathrm{sgn}(\beta)\left(\frac{\alpha^2}{2 \beta} -M\right) \right] \, .
\end{align*}
Next, we insert this approximation in Eq.~\eqref{eqA:stepEta}. While this step is only justified for small values of $\kappa$, it becomes exact in the limit $\kappa \rightarrow 0^+$, for which the heat-kernel regulator affects solely large energy solutions \footnote{To proof this statement, we use  the identity 
$\lim_{\kappa \rightarrow 0^+} \left[ \sum_{n=1}^\infty (\kappa n^{-1})^m \mathrm{e}^{ -\kappa c n } \right] =0$ for $m=1,2,3,\ldots$}.  
Using additionally the geometric series allows us to recast  Eq.~\eqref{eqA:stepEta}
\begin{align}
\eta_{_H}(\kappa)/n_\mathrm{0}\approx& \ 2s\;e^{-\kappa\sign (\beta) \left(\frac{\alpha^2}{2\beta}-M\right)}\sinh\left(\kappa \beta/2\right)\nonumber\\
	&\times \left[\frac{1}{1-e^{-\kappa\vert\beta\vert}}-1\right]+c \, e^{-\kappa \vert E_\mathrm{0}\vert}\ . \nonumber
\end{align}
The spectral asymmetry is defined as the analytic continuation for $\kappa\rightarrow 0^+$, this means $\eta_{_H}=\lim_{\kappa\rightarrow 0^+}\eta_{_H}(\kappa)$, resulting in
\begin{align}\label{eqA:finalEta}
\eta_{_H}= n_\mathrm{0} \, \text{sgn}\left( \mathrm{e}H\right)\left[\text{sgn}\left(M-\beta/2\right)+\text{sgn}\left(B\right)\right]. 
\end{align}
Equipped with Eqs.~\eqref{eqA:bareCharge} and~\eqref{eqA:finalEta}, we are finally in the position to calculate $j^0_\mathrm{bulk}(\mu,H)$. Let us first focus on the case $\mu=E_\mathrm{z}$ characterizing the ground state. Here, the bulk charge density  is determined by $j^0_\mathrm{bulk}(\mu=E_\mathrm{z},H) =-\mathrm{e} \langle\text{vac}\vert N \vert\text{vac}\rangle/S$, where $\vert\mathrm{vac}\rangle=\Pi_{n,k_\mathrm{x}}d_{n,k_\mathrm{x}}\vert 0\rangle$. Since the operators are normally ordered with respect to $E_\mathrm{z}$, $N_0\vert \text{vac}\rangle=0$. Thus in the ground state, the bulk charge carrier density at  $\mu=E_\mathrm{z}$ is given by
\begin{align*}
	&j^0_\mathrm{bulk}(\mu=E_\mathrm{z}, H )=\kappa_{_\mathrm{QAH}} H = \dfrac{\mathrm{e}}{2S} \eta_\mathrm{H} \ ,\\
	&\kappa_{_\mathrm{QAH}} =  \frac{\mathrm{e}^2}{2 \mathrm{h}}\left[\text{sgn}\left(M-\beta/2\right)+\text{sgn}\left(B\right)\right].
\end{align*}
This demonstrates that the Hall conductivity in the ground state is solely determined by $\eta_{_H}$ which is a signature of the parity anomaly \cite{Niemi83,Boyanovsky86}.
In contrast to the half-quantized Hall conductivity, obtained for a massive two-dimensional Dirac operator \cite{Niemi83,Schakel91}, we find that the effective mass parameter $B k^2$ takes the role of a regulator at high energies, resulting in the required integer quantization of the Hall conductivity \cite{Haldane88}. The asymmetry of the entire spectrum acts as if effectively a partner of the zeroth LL exists at large energies. 
Furthermore, Eq.~\eqref{eqA:finalEta} reveals that the spectral asymmetry vanishes when the $n=0$ LL crosses $E_\mathrm{z}$. This corresponds to the critical field $H_\mathrm{crit}$ found in App.~A, at which the LL spectrum loses all information on the band inversion. 
  
Now that we have determined $j^0_\mathrm{bulk}$ at $\mu=E_z$, we can compute the bulk charge carrier density for arbitrary $\mu$ [cf. Eq.~\eqref{eqA:fullParticle}]: 
\begin{align} 
j^0_\mathrm{bulk}(\mu,H)&=-\frac{\mathrm{e}}{S}\langle \Phi(\mu) \vert N  \vert \Phi(\mu)\rangle \nonumber\\
&= -\frac{\mathrm{e}}{S}\langle \Phi(\mu) \vert N_0  \vert \Phi(\mu)\rangle + \kappa_{_\mathrm{QAH}} H \, , \label{Nmuarb}
\end{align}
where $|\Phi(\mu)\rangle$ denotes a many-particle state for which all states are filled up to $\mu$, and $N_0$ is given by Eq.~\eqref{eqA:bareCharge}. This calculation is in general straightforward. The main difficulty lies in the fact that the $n=0$ LL can be either part of the valence or the conduction band, requiring a careful case analysis. To understand, in principle, how to explicitly evaluate Eq.~\eqref{Nmuarb}, let us consider the case for which $M/B>0$, $H<H_\mathrm{crit}$, and the $n=0$ LL is part of the valence band. In this case the ground state includes the $n=0$ LL:
\begin{align}
	\vert\mathrm{vac}\rangle=\prod_{n=0}^\infty \prod_{k_\mathrm{x}}d_{n,k_\mathrm{x}}\vert 0\rangle \, . \nonumber
\end{align}
With this state as the reference, we define the state for which all conduction band LLs with $n \leq N_\mathrm{max}$ are filled as
\begin{align}
	\vert \Phi( \mu_{N_\mathrm{max}}^{}) \rangle =\prod_{n=1}^{N_\mathrm{max}} \prod_{k_\mathrm{x}} b_{k_\mathrm{x},n}^\dagger \vert \mathrm{vac} \rangle \, , \nonumber
\end{align}
where $\mu_{N_\mathrm{max}}^{}$ defines the chemical potential between the two adjacent conduction band LLs with LL indices $N_\mathrm{max}$ and $N_\mathrm{max}+1$. To determine the charge carrier density for the given state, we still have to  evaluate the first term in Eq.~\eqref{Nmuarb}:
\begin{align}
	-\frac{\mathrm{e}}{S} \langle \Phi( \mu_{N_\mathrm{max}}^{}) \vert & N_0  \vert \Phi( \mu_{N_\mathrm{max}}^{}) \rangle  =  -\frac{\mathrm{e}}{S} n_0 N_\mathrm{max} \nonumber \\
	& = - \frac{ \mathrm{e}^2 } {h} \sum_{n=1}^{\infty}\theta(\mu_{N_\mathrm{max}}^{}-E_n^+) |H| \, . \nonumber
\end{align}
The recasting, as it is performed in the last step, will be necessary for writing our final result for an arbitrary $\mu$ in a simple and intuitive form later on. 

In an analogous fashion, we can determine the charge density for $\mu<E_z$. In this case, the first $N_\mathrm{max}$ LLs, including in particular the $n=0$ LL, are unoccupied:
\begin{align}
	\vert \Phi( \mu_{N_\mathrm{max}}^{}) \rangle =\prod_{n=0}^{N_\mathrm{max}} \prod_{k_\mathrm{x}} d_{k_\mathrm{x},n}^\dagger \vert \mathrm{vac} \rangle \, . \nonumber
\end{align}
Inserting this state into Eq.~\eqref{Nmuarb}, we again evaluate and recast its first term in the following way:
\begin{align}
	-\frac{\mathrm{e}}{S} \langle \Phi( \mu_{N_\mathrm{max}}^{}) \vert & N_0  \vert \Phi( \mu_{N_\mathrm{max}}^{}) \rangle =  \frac{\mathrm{e}}{S} n_0 N_\mathrm{max} \nonumber \\
	& = \frac{ \mathrm{e}^2 } {h} \sum_{n=0}^{N_\mathrm{max}}\theta(- \mu_{N_\mathrm{max}}^{}+E_n^-) |H| \, . \nonumber
\end{align}
We can repeat these steps for arbitrary $M,B$ and $H$ and arrive finally at
\begin{align*}
-\frac{\mathrm{e}}{S}\langle \Phi(\mu) \vert &N_0  \vert \Phi(\mu)\rangle 
  = \bigg\{-\kappa_{_\mathrm{QH}}^0  \Theta  \left ( \vert \mu \vert \! - \! \left \vert E_\mathrm{0}  \right \vert  \right) \\
& \  -  \sum_{\substack{n =1 \\ \, r=\pm }}^{\infty}  r \kappa_{_\mathrm{QH}}   \Theta \left[ r (\mu  -  E_n^r ) \right]\bigg\}H \, , 
\end{align*}
where $\kappa^0_{_\mathrm{QH}}$ and $\kappa_{_\mathrm{QH}}$ are defined in Eq.~(4) of our main text.

Our results can be easily extended to include a Zeeman Hamiltonian $\mathcal{H}_\mathrm{z}=\sigma_\mathrm{z} g_\mathrm{z} H$, or an additional exchange Hamiltonian $\mathcal{H}_\mathrm{ex}=\sigma_\mathrm{z} G_{\mathrm{ex}}(H)$ \cite{Liu08}. We only need to apply the replacement $M\rightarrow M+g_\mathrm{z} H+G_\mathrm{ex}(H)$. Since the extension to broken particle-hole symmetry is more tedious, further details will be given in Ref.~\cite{Boettcher18}, where it will be shown that our results are in general unaltered by the $D$-parameter \footnote{One only needs to replace $\mu \rightarrow \mu + D/l_H^2$ showing that the $D$-parameter acts as a chemical potential in magnetic fields.}.

Finally, let us comment on the effective bulk Lagrangian $\mathcal{L}_\mathrm{eff}^\mathrm{bulk}(\mu,H)$ characterizing the response of our system to a small perturbing field $a_\mu$ on top of an underlying background field $H$.  According to Eq.~\eqref{Nmuarb}, this small perturbation induces an additional bulk charge carrier density $j_\mathrm{ind}^0=\sigma_\mathrm{xy}\,\nabla \times \mathbf{a}$ on top of $j^0_\mathrm{bulk}$. One can then deduce the missing two spatial components of the induced three current $j^\mu_\mathrm{ind}$ by the requirement of Lorentz covariance \cite{Niemi83}:
\begin{align*}
j^\mu_\mathrm{ind}(\mu)= \sigma_{\mathrm{xy}}(\mu,H) \epsilon^{\mu\nu\rho}\partial_\nu a_\rho\ \, .
\end{align*}
Finally, the effective bulk Lagrangian follows from the fact that $j^\mu_\mathrm{ind}(\mu) = \delta \mathcal{S}_\mathrm{eff}^\mathrm{bulk}(\mu,H)/\delta a_\mu$, where the induced effective action is given by $\mathcal{S}_\mathrm{eff}^\mathrm{bulk}=\int \! \mathrm{d}^3 x \, \mathcal{L}_\mathrm{eff}^\mathrm{bulk}$.

\subsection*{QAH vs. QH response}
In the following, we elaborate further on the interpretation of Eq.~(4) of the main text [equivalently, Eq.~\eqref{Nmuarb}] and clarify the role of the Onsager relation. For simplicity, we assume $D=0$ and the electron charge $\mathrm{e}>0$.

\begin{figure}[b!]
\includegraphics[width=0.99\columnwidth]{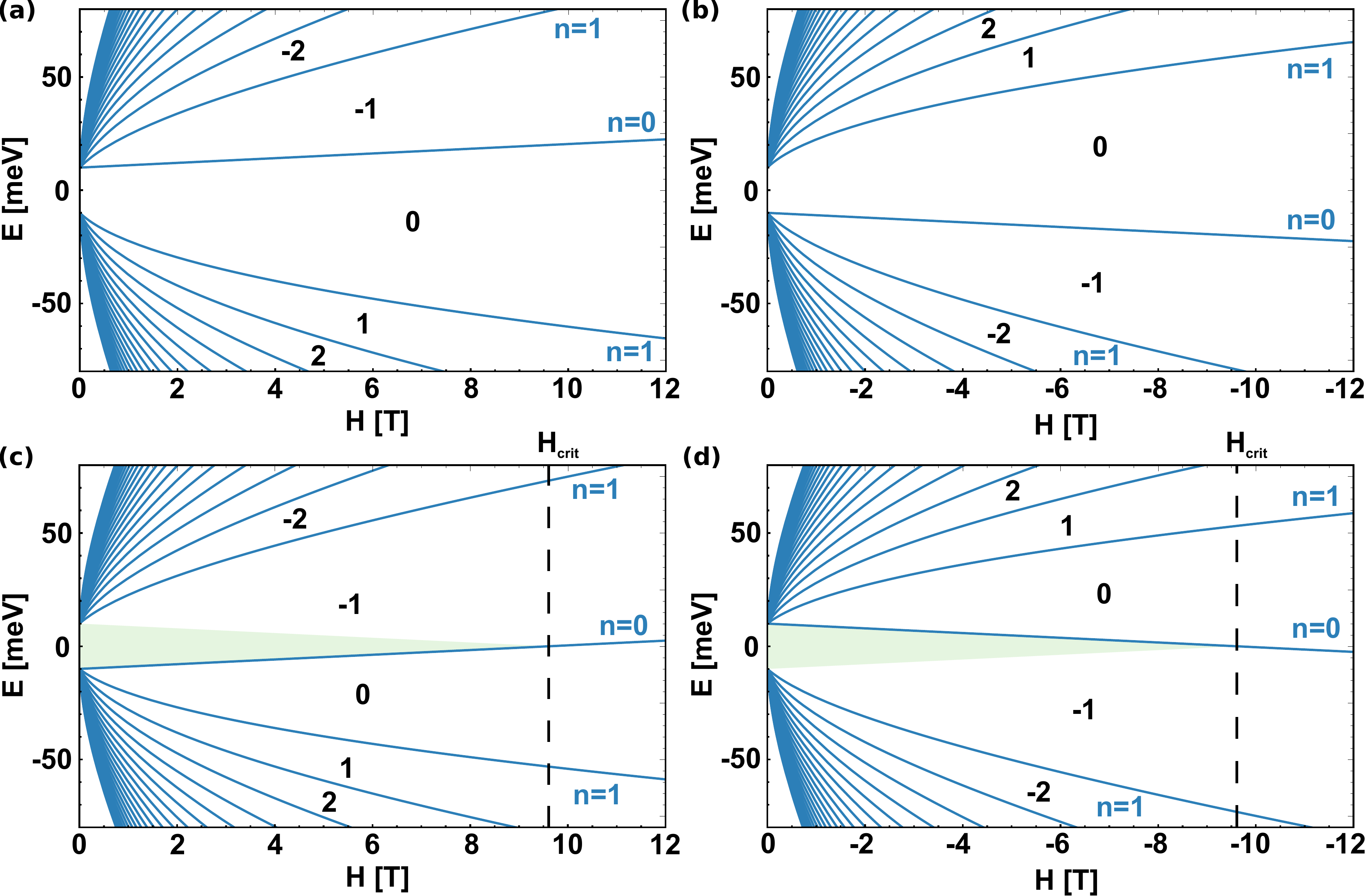}
\caption{\label{figA:GraphicalChernNumber}Evolution of bulk LL energies is depicted as function of orbital field (LL fan). For (a) and (c) H>0, while for (b) and (d) H<0.  Black numbers indicate the total Hall conductivity $\sigma_\mathrm{xy}\,[\mathrm{e}^2/\mathrm{h}]$ in the respective region. The Hall conductivity changes only if a LL (blue line) is crossed. For (a) and (b)  we use ${M=+10\, \text{meV}}$, $B=-685\, \text{meVnm}^2$, ${D=0\, \text{meVnm}^2}$, and $A=365\,\text{meVnm}$, implying that $\kappa_{_\mathrm{QAH}}=0$. For (c) and (d) the same parameters as in (a) and (b) are used except for $M= -10 \, \text{meV}$, resulting in $\kappa_{_\mathrm{QAH}}=-1$ for $H<H_\mathrm{crit}$. If below this critical field the chemical potential is placed such that $\vert \mu \vert < \vert E_0 \vert$ (depicted by green shaded area), the Hall conductivity remains invariant for \mbox{$H \rightarrow -H$}. For clarity, the $n=0$ LL and the first pair of LLs with $n=1$ is explicitly marked in all plots.}
\end{figure}

Let us first focus on the trivial case $M/B<0$, implying that only $\kappa_{_\mathrm{QH}}^0$ and $\kappa_{_\mathrm{QH}}$ contribute to the total Hall conductivity $\sigma_\mathrm{xy}$ since $\kappa_{_\mathrm{QAH}}=0$. In this case, the evolution of the LL energies as a function of the orbital field, given by Eqs.~\eqref{eqAp:nLLenergies} and \eqref{eqAp:0LLenergies}, is shown in Figs.~\ref{figA:GraphicalChernNumber} (a) and (b) for positive and negative orbital fields, respectively.
For $n\geq1$, all LLs come in pairs $E_{n\neq0}^\pm$ indicating that every conduction band LL has a partner in the valence band.  According to Eq.~(4c), each of these valence (conduction) band LL contributes $\sigma_\mathrm{xy}=+\mathrm{e}^2/\mathrm{h}$ ($\sigma_\mathrm{xy}=-\mathrm{e}^2/\mathrm{h}$) to the total Hall conductivity for $H>0$. For $H<0$, they contribute with the opposite  sign. We identified this $\mathrm{sgn}(\mathrm{e}H)$ dependence in the main text as the characteristic feature of a conventional QH / LL response.
The $\mathrm{sgn}(\mathrm{e}H)$-dependence also holds for the $n=0$ LL, however with the peculiar difference that it lacks a partner \cite{Haldane88}. This implies that there is only one $n=0$ LL which is either part of the valence or of the conduction band, as can be seen by comparing Fig.~\ref{figA:GraphicalChernNumber} (a) with (b). This asymmetry is reflected by Eq.~(4b) which shows the characteristic $\mathrm{sgn}(\mathrm{e}H)$ dependence, but also contains information about the absence of a partner. Finally, let us remark that one would need to shift the chemical potential from the conduction into the valence band ($\mu \rightarrow - \mu$)  to observe for $H>0$ and $H<0$ the same sign of the Hall conductivity.

Let us now focus on the non-trivial case where \mbox{$M/B>0$}. Here, all statements made for  $\kappa_{_\mathrm{QH}}$ and $\kappa_{_\mathrm{QH}}^0$ are still valid but, in contrast to the trivial case, $\kappa_{_\mathrm{QAH}}$ contributes now additionally to the total Hall conductivity.
This scenario is shown in Figs.~\ref{figA:GraphicalChernNumber} (c) and (d). Since $\kappa_{_\mathrm{QAH}} \neq 0$ for $H<H_\mathrm{crit}$, there is a range of chemical potentials for which the Hall conductivity does not change its sign for $H \rightarrow -H$. To be precise, this regime is given by $\vert \mu \vert < \vert E_0 \vert$, meaning that the chemical potential must be placed in the Dirac mass gap (indicated by the green shaded area in Fig.~\ref{figA:GraphicalChernNumber}).
Physically, this terms implies that the conventional Landau level physics ($\kappa_{_\mathrm{QH}}$ and $\kappa_{_\mathrm{QH}}^0$) comes on top of an overall, underlying Chern number $\kappa_{_\mathrm{QAH}}$. According to Eq.~\eqref{eqA:finalEta}, this Chern number is related to the spectral asymmetry and  shifts the Hall conductivity such that it becomes $\sigma_\mathrm{xy}=-\mathrm{e}^2/\mathrm{h}$ (for $M,B<0$) in the Dirac mass gap.
We identified this regime in the main text as the hallmark of the QAH response which remains encoded in orbital fields. Note that this property is in accordance  with Streda's formula but implies that the Onsager relation, $\sigma_\mathrm{xy}(-H)=-\sigma_\mathrm{xy}(H)$, is violated in the Dirac mass gap due to the parity anomaly. To be even more precise, we refer to a violation of the Onsager relation in the sense that in the Dirac mass gap $\sigma_\mathrm{xy} (-H) =\sigma_\mathrm{xy}(H)$. 


\section{Appendix C: Parity symmetry}
In this appendix, we systematically analyze in which way the building blocks of the Chern Hamiltonian in Eq.~\eqref{eqA:ChernHamiltonAp} change under parity transformation. In (2+1)D, a parity transformation $\mathcal{P}$ is defined via \cite{Deser82}
\begin{align}
r&= (t,x,y) \overset{\mathcal{P}}{\rightarrow} (t,-x,y) = \tilde{r}, \nonumber\\
k&=(\omega,k_\mathrm{x},k_\mathrm{y}) \overset{\mathcal{P}}{\rightarrow} (\omega,-k_\mathrm{x},k_\mathrm{y}) = \tilde{k}.  \nonumber 
\end{align}
From this, we can deduce that the fermionic spinor operators transform under parity as \cite{Deser82}
\begin{align}
\psi(r) \overset{\mathcal{P}}{\rightarrow} \sigma_\mathrm{y} \psi(\tilde{r}). \nonumber
\end{align}
Thus, it follows that the building blocks of the Chern Hamiltonian transform under parity via
\begin{align*}
& \psi^\dagger(r)  M \, \sigma_\mathrm{z}  \psi(r)   & \overset{\mathcal{P}}{\rightarrow} &      &-     \psi^\dagger(r) M \, \sigma_\mathrm{z} \psi(r), &  \\
  & \psi^\dagger(r) Bk^2 \, \sigma_\mathrm{z}  \psi(r)   & \overset{\mathcal{P}}{\rightarrow}&     &-    \psi^\dagger(r) Bk^2 \, \sigma_\mathrm{z} \psi(r),  & \\
 & \psi^\dagger(r) Dk^2 \, \sigma_\mathrm{0}  \psi(r)  &  \overset{\mathcal{P}}{\rightarrow} &     &  \psi^\dagger(r)  Dk^2 \, \sigma_\mathrm{0} \psi(r), & \\
  & \psi^\dagger(r)  A k_\mathrm{x} \, \sigma_\mathrm{x}   \psi(r)  &  \overset{\mathcal{P}}{\rightarrow} &     &  \psi^\dagger(r) A k_\mathrm{x} \, \sigma_\mathrm{x}  \psi(r), & \\
  & \psi^\dagger(r) A k_\mathrm{y} \, \sigma_\mathrm{y}  \psi(r)  &  \overset{\mathcal{P}}{\rightarrow}  &   &  \psi^\dagger(r) A k_\mathrm{y} \, \sigma_\mathrm{y}  \psi(r) & \ .
\end{align*}
Since the Dirac mass operator $\psi^\dagger(r) M \sigma_\mathrm{z}  \psi(r) $, as well as the effective  mass operator $ \psi^\dagger (r)Bk^2 \sigma_\mathrm{z}  \psi(r) $ change sign under $\mathcal{P}$, they break parity symmetry explicitly.
 
In Eq.~(3) of the main text, we constructed an effective action for a Chern insulator in terms of a small perturbing vector potential $a_\mu$. This was done by effectively integrating out the fermionic sector. Using this procedure, a Chern-Simons term of odd parity was induced \cite{Deser82}:
\begin{align}
\epsilon^{\mu \nu \rho} a_\mu \partial_\nu a_\rho \quad \overset{\mathcal{P}}{\rightarrow} \quad -  \epsilon^{\mu \nu \rho} a_\mu \partial_\nu a_\rho. \nonumber
\end{align}
For zero orbital field, this parity-breaking Chern-Simons term arose from the parity breaking mass terms $M$ and $B$, as discussed above. Hence for $H=0$, we expect that the Chern-Simons level is exclusively a function of these parameters. In particular, we find that
\begin{align}
\mathcal{C}_\mathrm{QAH}=  \left[\mathrm{sgn}(M)+\mathrm{sgn}(B)\right]/2 \ , \nonumber
\end{align}
which is in agreement with Ref.~\cite{Lu10}.


\section{Appendix D: Numerical approach}
This appendix gives more details on the numerical approach, which was employed to study  QAH insulators in  
time-dependent orbital fields $H(t)$. In particular, we show how we were able to visualize the charge pumping shown in Fig.~1 and 2 of the main text. We apply the Peierls substitution in the Landau gauge to introduce the orbital field $H(t)$. This implies that $k_\mathrm{x}$ is a good quantum number and, therefore, enables us to write the Hamiltonian and its corresponding Hilbert space as a direct sum:
\begin{align}
\mathcal{H}(t) = \bigoplus_{k_\mathrm{x}} \mathcal{H}_{k_\mathrm{x}}(t) \ . \nonumber
\end{align}
The numerical simulations can be therefore carried out  on each Hilbert subspace $\mathcal{H}_{k_\mathrm{x}}$ separately.

At the initial time $t_\mathrm{i}$, the eigenstates of Eq.~\eqref{eqA:ChernHamiltonAp} are solutions of the time-independent Schr\"{o}dinger equation:
\begin{align}
\mathcal{H}(t_\mathrm{i}) \vert\psi_{j,k_\mathrm{x}} (t_\mathrm{i})\rangle =E_{j,k_\mathrm{x}}(t_\mathrm{i}) \vert\psi_{j,k_\mathrm{x}} (t_\mathrm{i})\rangle \ . \nonumber
\end{align}
The orbital field is now increased as a function of time and the evolution of all eigenstates is traced via the time-dependent Schr\"{o}dinger equation,
\begin{align}
\im \hbar \partial_t \vert\psi_{j,k_\mathrm{x}} (t)\rangle = \mathcal{H}(t) \vert\psi_{j,k_\mathrm{x}} (t)\rangle \ .  \nonumber
\end{align}
We compute the time-evolution of eigenstates numerically, using an iterative procedure:  
\begin{align}
 \vert\psi_{j,k_\mathrm{x}} (t+ \Delta t)\rangle & = \mathrm{e}^{-\im \mathcal{H}(t) \Delta t/\hbar}    \vert\psi_{j,k_\mathrm{x}} (t)\rangle   \nonumber  \\
& = U(t+\Delta t,t)  \vert\psi_{j,k_\mathrm{x}} (t)\rangle \ ,  \nonumber
\end{align}
where $U(t_2,t_1)$ denotes the unitary time evolution of each state from $t_1\!\rightarrow\!t_2$ and $\Delta t$ has to be chosen small enough to ensure convergence. After the time $t$, we obtain
\begin{align} \label{timeevoloftates}
 \vert\psi_{j,k_\mathrm{x}} (t)\rangle =U(t,t_\mathrm{i})  \vert\psi_{j,k_\mathrm{x}} (t_\mathrm{i})\rangle \ .  
\end{align}
\noindent
We apply this iterative procedure to analyze the evolution of the following non-interacting, many-particle state in orbital fields, where all states are filled up to the chemical potential $\mu$:
\begin{align*}
|\Phi(\mu,t=t_\mathrm{i})\rangle=\!\!\!\prod_{\substack{j \leq j_\mathrm{max} \\ k_\mathrm{x} \leq k_\mathrm{max}}} \! \! \vert\psi_{j,k_\mathrm{x}} (t_\mathrm{i})\rangle \, ,
\end{align*}
where $j_\mathrm{max}$ and $k_\mathrm{max}$ are determined by a given $\mu$.
 
Now, tracing these initially filled states via Eq.~\eqref{timeevoloftates} 
enables us to determine two characteristic, time-dependent quantities.
Firstly, we can compute the induced charge density distribution, which was used in Fig.~2 of the main text, to study charge flow in QAH insulators:
\begin{align}\label{eqA:indCharge}
j^0_\mathrm{ind}(\mathbf{x},t) =  -\mathrm{e}\!\!\!\!\!\sum_{\substack{ k_\mathrm{x}\leq k_\mathrm{max} \\  j \leq j_\mathrm{max}}} \!\!\!\!\! \psi_{j,k_\mathrm{x}}^\dagger (\mathbf{x},t) \  \psi_{j,k_\mathrm{x}} (\mathbf{x},t) -j^0_\mathrm{back} \, ,
\end{align}
where $j^0_\mathrm{back}$ ensures $j^0_\mathrm{ind}(\mathbf{x},t_i)\!=\!0$. Secondly, we can identify the states which are responsible for this charge flow. Therefore, we trace the filling probabilities of each instantaneous eigenstate  at time $t$, in the time-independent Schr\"{o}dinger equation $\mathcal{H}(t)\vert\phi_{i,k_\mathrm{x}}(t)\rangle=~E_{i,k_\mathrm{x}}(t)\vert\phi_{i,k_\mathrm{x}}(t)\rangle$. Note that here $t$ is not a dynamical variable, defining the time evolution of states as in Eq.~\eqref{timeevoloftates}, but rather parametrizes the eigensystem of the Hamiltonian at time $t$. In particular, the occupation probability of an eigenstate $\vert\phi_{i,k_\mathrm{x}}(t)\rangle$, depicted in Fig.~1 of the main text, is given by:
\begin{align} \label{probs}
P_{i,k_\mathrm{x}}(t)\!=\!\sum_{j \leq j_\mathrm{max}} \vert\langle\psi_{j,k_\mathrm{x}}(t)\vert\phi_{i,k_\mathrm{x}}(t)\rangle\vert^2 \, .
\end{align}
This quantity can be used to recast Eq.~\eqref{eqA:indCharge} to the form, shown in the main text:
\begin{align*}
j^0_\mathrm{ind}(\mathbf{x},t)=-\mathrm{e}\sum_{i,k_\mathrm{x}} P_{i,k_\mathrm{x}}(t)|\phi_{i,k_\mathrm{x}}(\mathbf{x},t)|^2 - j^0_\mathrm{back} \, .
\end{align*}
 
If our results are supposed to be experimentally accessible, the results shown in Fig.~1 and 2 of the main text should not depend on how fast we increase the orbital field.
In the numerical approach, we raise $H(t)$ within the time interval $[t_\mathrm{i}=0,t_\mathrm{f}]$ corresponding to a ramping speed 
\begin{figure}[t!]
\centering
\includegraphics[width=.95\columnwidth]{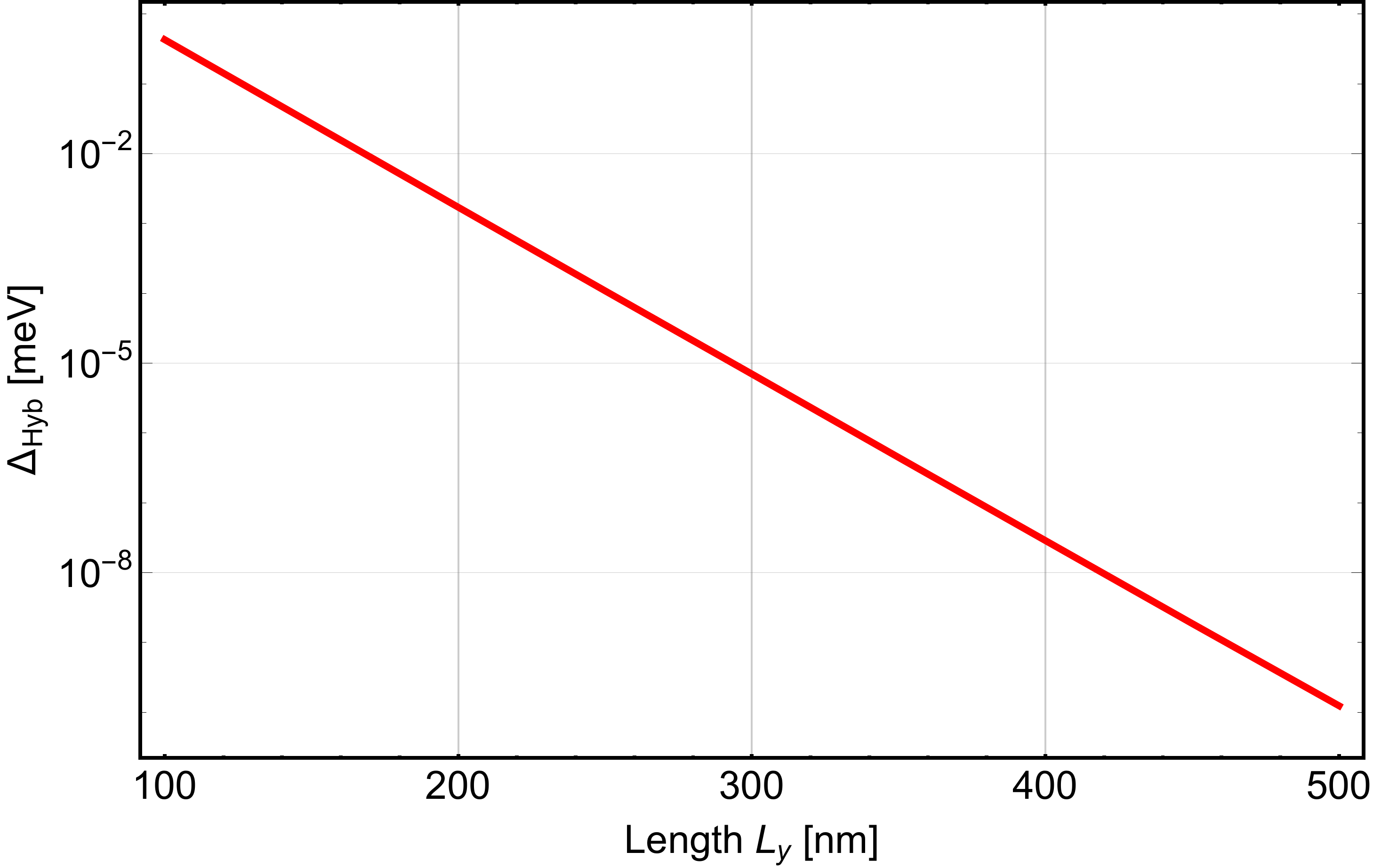}
\caption{\label{fig:hybGap} Finite size gap $\Delta_\mathrm{Hyb}(L_\mathrm{y})$ for $k_\mathrm{x}=0$ and $H_\mathrm{scat}<H\!<\!H_\mathrm{crit}$, forming between QAH edge states and the $n=0$ LL, as a function of the system length $L_\mathrm{y}$ (depicted in Log-Lin plot). Parameters are taken from the caption of Fig.~1 in the main text.  The linear behavior shows that $\Delta_\mathrm{Hyb}(L_\mathrm{y})$ drops exponentially with $L_\mathrm{y}$. Here, we find that $\Delta_\mathrm{0}=100 \,$meV, and $\lambda=0.05 \,\mathrm{nm}^{-1}$ (cf. Eq.~\eqref{eqA:finGap}). We obtain analogous results for $k_\mathrm{x} \neq 0$.}
\end{figure}
\begin{align} \label{rampingspeed}
v_\mathrm{ramp} = \dfrac{H_\mathrm{max}}{t_\mathrm{f}} \quad \mathrm{with} \quad H_\mathrm{max} = H \left( t_\mathrm{f} \right)    \ .
\end{align}
For a fixed  $H_\mathrm{max}$, different ramping speeds can be therefore simulated by varying $t_\mathrm{f}$.  Our results in Fig.~1 and 2 of the main text fulfill the following requirements: the  ramping time $t_\mathrm{f}$ and, therefore, the ramping speed $v_\mathrm{ramp}$ have to be chosen such that
\begin{align} \label{limits}
t_\mathrm{f}^\mathrm{min} \overset{(i)}{\ll} t_\mathrm{f} \overset{(ii)}{\ll} t_\mathrm{f}^\mathrm{max} \ ,
\end{align}
where $(ii)$ only needs to be fulfilled for $H\!>\!H_\mathrm{scat}$ [Fig.1~(c) of the main text].

$(i)$ The lower (upper) bound on $t_\mathrm{f}$ ($v_\mathrm{ramp}$) results from the fact that particles should not be excited between bulk bands. Therefore, $H(t)$ has to be increased on a time scale which is adiabatic with respect to any bulk energy gap $E_\mathrm{g}$. In particular, this implies that $t_\mathrm{f}^\mathrm{min} \ll t_\mathrm{f}$ with
\begin{align} \label{eqA:tmin}
t_\mathrm{f}^\mathrm{min} = \frac{\hbar}{E_\mathrm{g}} \, . 
\end{align}
In order to overcome $E_\mathrm{g}$, which can be on the order of a few tenth of meV, we would need to ramp up $H_\mathrm{max}$ (a few Tesla) on a very small time scale $t_\mathrm{f} \ll 10^{-13}$s. 

$(ii)$  The upper (lower) bound on $t_\mathrm{f}$ ($v_\mathrm{ramp}$) is caused by the fact that, for $H\!>\!H_\mathrm{scat}$, unoccupied QAH edge states and occupied bulk LLs form finite hybridization gaps $\Delta_\mathrm{Hyb}$ as the QAH edge states are lowered in energy with increasing $H(t)$ (cf.~Fig.~1(c) of the main text). Our goal is to ensure that the QAH edge states and all bulk LLs separately maintain their initial filling probabilities throughout this process. As a result, for $H\!>\!H_\mathrm{scat}$, we need to choose $t_\mathrm{f}$ such that we diabatically overcome $\Delta_\mathrm{Hyb}$. Diabatically means that neither the filling probabilities nor the local densities of the QAH edge states and the bulk LL wave functions change, if they pass each other with increasing orbital field. Analogous to Eq.~\eqref{eqA:tmin}, this implies that  $t_\mathrm{f} \ll t_\mathrm{f}^\mathrm{max}$ with
\begin{align*}
t_\mathrm{f}^\mathrm{max}  = \dfrac{\hbar}{\Delta_\mathrm{Hyb}} \, .
\end{align*} 
If $\Delta_\mathrm{Hyb}$ are finite size gaps (cf. App.~A.2), satisfying
\begin{align} \label{eqA:finGap}
\Delta_\mathrm{Hyb}(L_\mathrm{y}) \  = \Delta_\mathrm{0} \ \mathrm{e}^{- \lambda \, L_y} \quad \mathrm{with} \quad \lambda >0 \ ,
\end{align}
time-scales, which are experimentally possible to reach, become accessible since $t_\mathrm{f}^\mathrm{max}$ increases exponentially. As shown in Fig.~\ref{fig:hybGap}, we find for $H_\mathrm{scat}\!<H\!<\!H_\mathrm{crit}$ and $k_\mathrm{x}\!=\!0$ (same holds for $k_\mathrm{x}\!\neq\!0$) that energy gaps forming between QAH edge states and the $n\!=\!0$ LL
are exponentially small. In a typical macroscopic Hall bar \cite{Budewitz17}, the system length can be on the order of $L_\mathrm{y}\!\approx\! 10 \, \mu \mathrm{m}$ \cite{Budewitz17}, which  implies that $t_\mathrm{f}^\mathrm{max}$  can be  approximately infinite compared to  all other experimental time scales. It is therefore plausible that even for $H>H_\mathrm{scat}$ the QAH charge pumping could be experimentally observable in macroscopically large systems provided that scattering between the QH and the QAH edge states can be strongly suppressed. However, in a conventional device elastic and inelastic scattering events between the QAH and the QH edge states cause relaxation of the charge inversion which ultimately leads to a transition to region \RM{2} indicated in Fig.~3 of the main text.

\section{Appendix E: Scattering between QH and QAH edge states}

\begin{figure}[!t]
\centering
\includegraphics[width=.9\columnwidth]{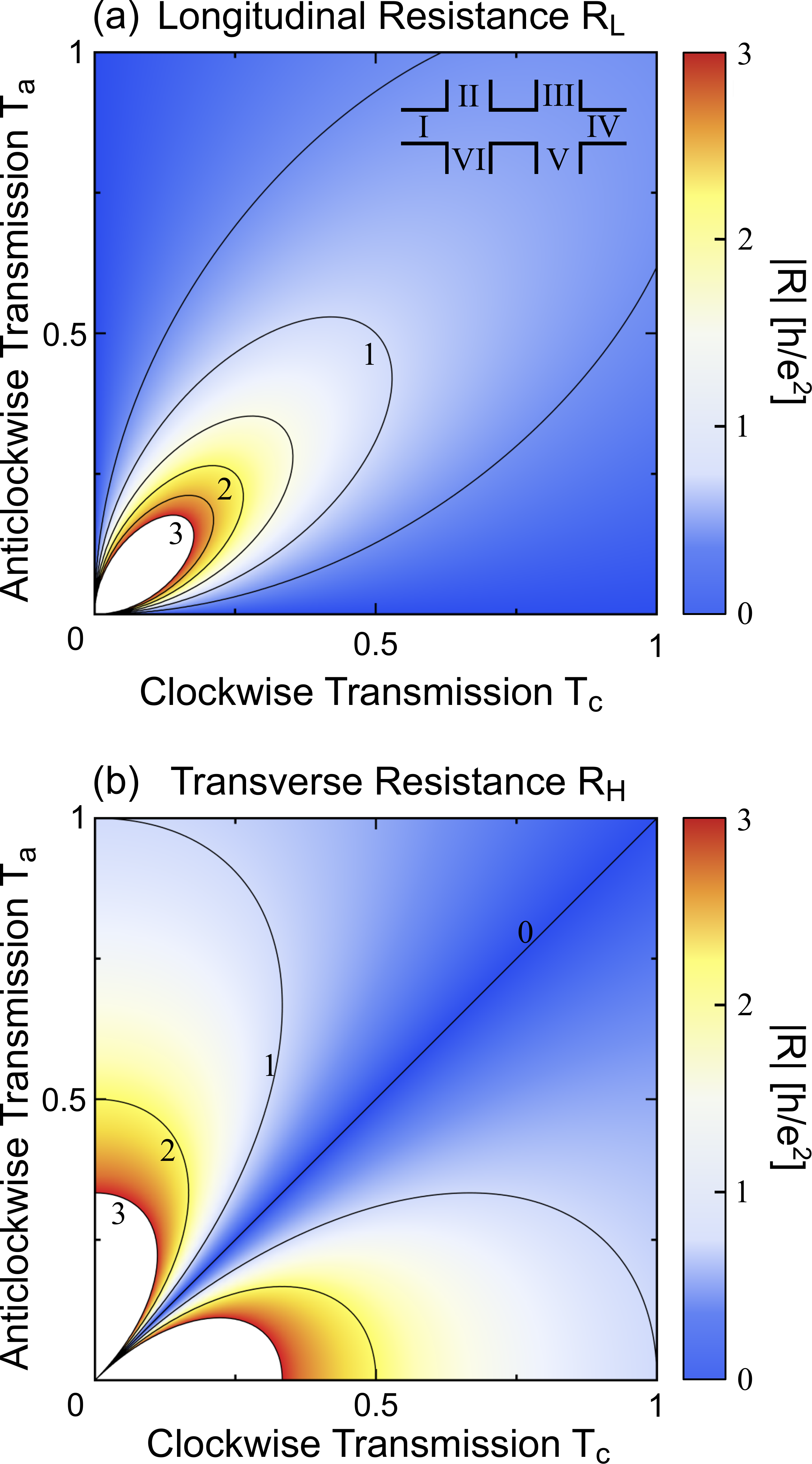}
\caption{We map out the full phase space for (a) longitudinal  and (b)  (transverse) Hall resistance with $T_{12},T_{21}\leq 1$. The underlying six-terminal Hall bar is schematically depicted in the inset of (a).  Current flows between contact $1$ and $4$. The color code highlights the absolute value of resistance with white indicating that the resistance is out of scale. Contour lines highlight in (a)  $R_\mathrm{L}=0.5,1,1.5,2,2.5,3$  and in (b) $R_\mathrm{H}=0,1,2,3$ (given in units of $\mathrm{h}/\mathrm{e}^2$).  \label{FigSupp:BuettikerHallMeasurement}}
\end{figure}

In the main text, we discussed transport signatures of counterpropagating QAH and QH edge states (cf. Fig. 3 regime \RM{2}) within the Landauer-B\"uttiker approach. We proposed that transmission probabilities between adjacent voltage probes can differ for QH and QAH edge states.
Within this formalism \cite{Buettiker88,Wang13}, the current in the $i$-th contact is given by
\begin{align} \label{eqA:Landauer}
I_i = -\frac{\mathrm{e}}{\mathrm{h}}\sum_{j=1}^N \left[ T_{ij} \mu_j - T_{ji} \mu_i \right],
\end{align}
where $T_{ij}$ is the transmission probability from contact $j$ to contact $i$, $N$ is the number of contacts, and $\mu_i$ is the local chemical potential in the $i$-th probe. In a typical Hall bar with $N=6$, as schematically shown in the inset of Fig.~\ref{FigSupp:BuettikerHallMeasurement}(a), contact 1 and 4 act as source and drain of current with $I_1=-I_4$ and the four remaining contacts are voltage probes with $I_j = 0$. The Hall resistance is defined by $R_\mathrm{H} = R_{26} / I_1$; the longitudinal resistance is given by $R_\mathrm{L} = R_{23} / I_1$. To discuss transport signatures of our system, let us first assume that the transmission probabilities in clockwise and anticlockwise direction are determined by $T_{i+1,i}=T_\mathrm{c}$ and $T_{i,i+1}=T_\mathrm{a}$, respectively. Solving the linear system, given by Eq.~\eqref{eqA:Landauer}, leads to the following analytic expressions for Hall and longitudinal resistance,
\begin{align}
R_\mathrm{H} &=\frac{\mathrm{h}}{\mathrm{e}^2} \frac{T_\mathrm{c}-T_\mathrm{a}}{T_\mathrm{c}^2-T_\mathrm{a} T_\mathrm{c} +T_\mathrm{a}^2},\\
R_\mathrm{L}&=\frac{\mathrm{h}}{\mathrm{e}^2} \frac{T_\mathrm{c}T_\mathrm{a}}{T_\mathrm{c}^3+T_\mathrm{a}^3}.
\end{align}
In Fig.~\ref{FigSupp:BuettikerHallMeasurement}, we map out the full parameter space for Hall and longitudinal resistance, taking $ T_\mathrm{c},T_\mathrm{a} \leq 1$. If $T_\mathrm{c}=T_\mathrm{a}=1$, we reach the characteristic values of a QSH phase, $R_\mathrm{H}=0$ and $R_\mathrm{L}= \mathrm{h} / 2 \mathrm{e}^2$ \cite{Konig08}. If $T_\mathrm{c}=1$ and $T_\mathrm{a}=0$ (or $T_\mathrm{c}=0$ and $T_\mathrm{a}=1$), the transport signature is equivalent to the one of  a single chiral mode.  As our system contains two counterpropagating edge states which are not protected by symmetry, realistic transmission probabilities can deviate from these limiting cases.
Nevertheless, resistance values in the vicinity of the contour lines depicted in Fig.~\ref{FigSupp:BuettikerHallMeasurement}, can be still close to quantized values for a large range of parameters. In addition, Fig.~\ref{FigSupp:BuettikerHallMeasurement}(b) demonstrates that small deviations from the symmetric case (i.e., from $T_c=T_a$) cause large deviations from $R_\mathrm{H}=0$, if $T_c, T_a \ll 1$.

Let us now investigate QH and QAH edge states in more detail. Since these states are localized on the same edge and scattering between them is allowed by symmetry, point-like impurities can already give rise to backscattering. Due to the required unitarity of the $S$-matrix, we find in this case that $T_\mathrm{c}=T_\mathrm{a}$, which both tend to  zero in the large system limit. However, any small difference between the two sides of the Hall bar can cause slight deviations from a perfect quantization, as indicated schematically by a noisy plateau in regime \RM{2} (cf. Fig.~3 in the main text). 

Another very prominent source for  backscattering are charge puddles \cite{Vayrynen13} constituting a major, if not the dominant, source for backscattering in HgTe based two-dimensional topological insulators \cite{Lunczer19}. The characteristic value of $R_\mathrm{L}=\mathrm{h}/2\mathrm{e}^2$ has been therefore only achieved in micro-structured Hall bars. Here, we want to focus however on large samples, where $L>n_\mathrm{p}^{-1/2}$ and $n_\mathrm{p}$ is the puddle density \cite{Vayrynen13}. In this limit, V\"{a}yrynen et al. \cite{Vayrynen13} showed that the bulk conductivity cannot be neglected if the system size exceeds the leakage length $L^\star=1/\sigma_\mathrm{B} \rho_\mathrm{e}$. Here, $\sigma_\mathrm{B}$ is the bulk conductivity and $\rho_\mathrm{e}$ is the edge resistivity.

\begin{figure}[!t]
\centering
\includegraphics[width=.95\columnwidth]{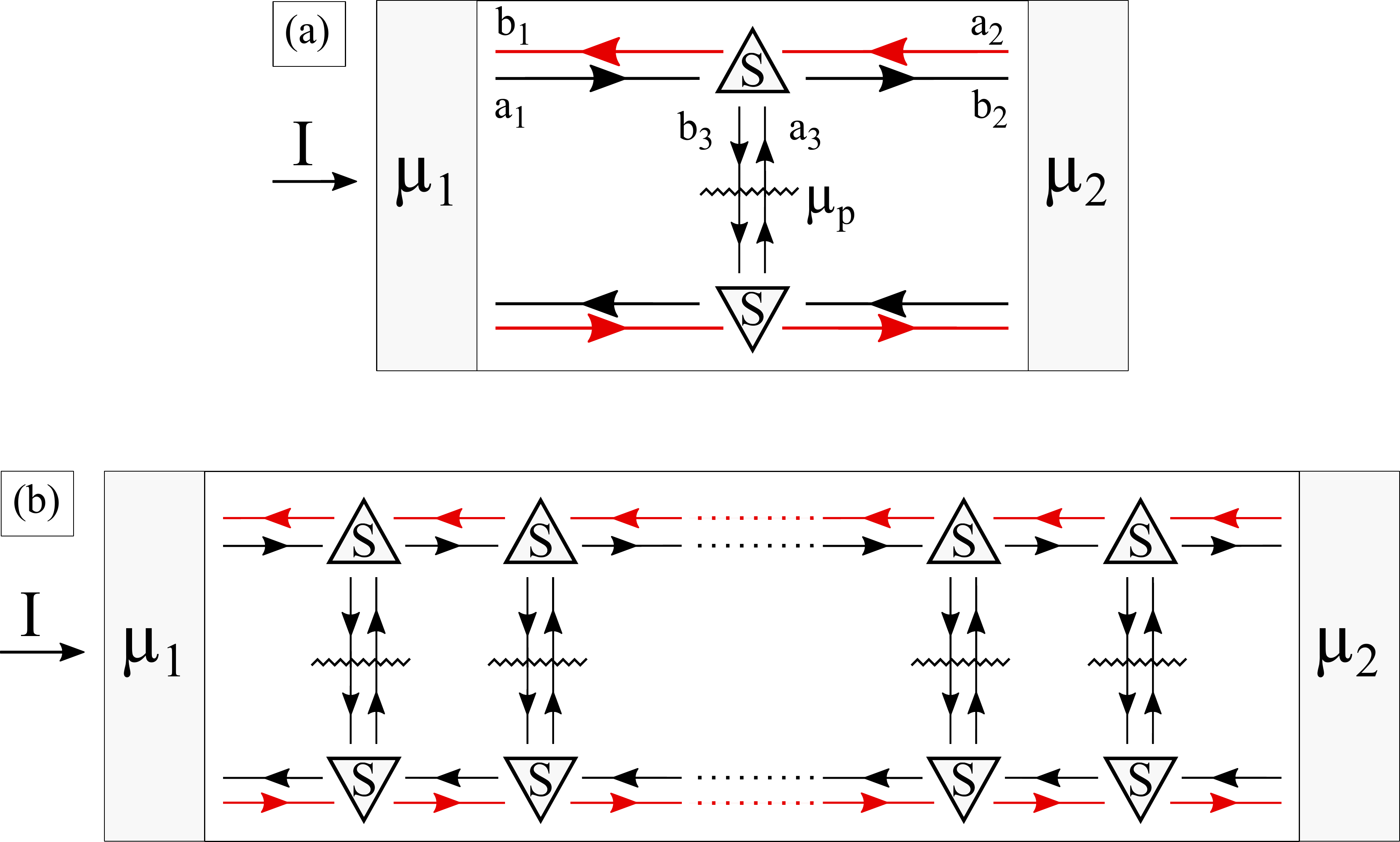}
\caption{Schematic two-terminal set-up, where counterpropagating QH and QAH edge states are depicted by red and black arrows at the top and the bottom edge. The triangle indicates a scatterer with scattering matrix $S$ that partially transmits particles coherently and partially scatters particles into a fictitious contact (wiggly line) with chemical potential $\mu_\mathrm{p}$. The fictious contact models an inelastisc scattering event connecting top and bottom edge. In (a), we consider a toy model with a single charge puddles whereas (b) generalizes the situation to the case of $N$ charge puddles. In (a), incoming $a_i$ and outgoing $b_i$ scattering states are labeled for the top edge.  \label{FigSupp:PuddleScattering}}
\end{figure}

If $L>L^\star$, the top and bottom edge can be connected via puddle-to-puddle hopping. To gain a better understanding of this situation, we study a toy model using the Landauer-B\"{u}ttiker approach. Here, the top and bottom edge are connected via a single charge puddle (in a realistic situation, an electron would need to hop multiple times between adjacent puddles to reach the other edge). This situation is schematically depicted in Fig.~\ref{FigSupp:PuddleScattering}(a). The scattering from the edge states onto the charge puddle can be described by the following $S$-matrix:
\begin{align}
S=\begin{pmatrix}
r_{11} & t_{12} & t_{13} \\
t_{21} & r_{22} & t_{23} \\
t_{31} & t_{32} & r_{33}
\end{pmatrix},
\end{align}
where $t_{ij}$ and $r_{ij}$ denote transmission and reflection amplitudes from the $j$-th incoming to the $i$-th outgoing scattering state, respectively. For the top edge, scattering states are labeled according to Fig.~\ref{FigSupp:PuddleScattering}(a). For the bottom edge, we assume the same type of scatterer but QH and QAH edge states switch their propagation direction. This model describes partially coherent transmission \cite{Buettiker86} of QH and QAH edge states ($t_{12}$ and $t_{21}$), where only a fraction of the current is transmitted onto the charge puddle ($t_{31}$ and $t_{32}$). Since charge puddles act like inelastic scatterers, they cause dephasing and can be therefore modeled as fictious voltage probes \cite{Buettiker88}. 

Since QH and QAH edge states have different spin character and wave function localization, the QH-to-puddle tunneling probability ($|t_{32}|^2$) can differ from the QAH-to-puddle tunneling probability ($|t_{31}|^2$). We start with a specific model to prove the possibility of asymmetric transmission probabilities $T_{ij}$. Choosing $r_{11}=r_{22}=t_{32}=t_{13}=0$, unitarity of the $S$-matrix requires that
\begin{align*}
T_\mathrm{QH}=1, \quad  \quad T_\mathrm{QAH}+T_\mathrm{p}=1, \quad  \quad R_\mathrm{p}+T_\mathrm{p}=1,
\end{align*}
where $T_\mathrm{QAH}=|t_{21}|^2$, $T_\mathrm{QH}=|t_{12}|^2$, $T_\mathrm{p} = |t_{31}|^2=|t_{23}|^2$, and $R_\mathrm{p}=|r_{33}|^2$.  Importantly, $T_\mathrm{p}$ denotes the transmission probability from the chiral QAH edge state to the charge puddle. Without loss of generality, we take $\mu_1 > \mu_2$ and $\mu_2=0$. The current into the puddle is therefore given by
\begin{align}
I_\mathrm{p}=-\frac{\mathrm{e}}{\mathrm{h}} \left[ \left(2-2 R_\mathrm{p}\right)  \mu_\mathrm{p}-T_\mathrm{p} \, \mu _1 \right].
\end{align}
With $I_\mathrm{p}=0$, it follows that $\mu_\mathrm{p} = \mu_1 /2$. The current which flows along the top edge into contact $2$ is given by
\begin{align}
I_2&=-\frac{\mathrm{e}}{\mathrm{h}} \left( T_\mathrm{QAH} \, \mu_1+T_\mathrm{p} \, \mu_\mathrm{p}\right)\\
&=-\frac{\mathrm{e}}{\mathrm{h}}\left(1-T_\mathrm{p}/2\right)\mu_1. \label{eqA:CurrentOnePuddle}
\end{align} 
We can identify $1-T_\mathrm{p}/2$ as an effective transmission coefficient between contact 1 and 2, i.e., $T_{21}$. Since $T_{12}=T_\mathrm{QH}=1$ and $T_{21}<1$ for $T_\mathrm{p} \neq 0$, we showed the possibility of having asymmetric transmission coefficients when top and bottom edge states are connected via puddle-to-puddle hopping.

As $T_\mathrm{p}$ is in general a small number, it is interesting to look at the case of many charge puddles. This situation is schematically illustrated in Fig.~\ref{FigSupp:PuddleScattering}(b). Following an analogous calculation, it is straightforward to generalize Eq.~\eqref{eqA:CurrentOnePuddle} to the situation of $N$ puddles:
\begin{align}\label{eqA:manyPuddles}
I_2=-\frac{\mathrm{e}}{\mathrm{h}}\frac{2-T_\mathrm{p}}{2+(N-1)T_\mathrm{p}}\mu_1.
\end{align}
In conclusion, this shows that, for $L>L^\star$, it is possible to find peculiar values for $R_\mathrm{H}$ and $R_\mathrm{L}$ in magnetotransport experiments. In particular, the Hall conductivity can deviate from zero. Intriguingly, it is even possible to measure a Hall resistance close to $R_\mathrm{H}=\mathrm{h}/\mathrm{e}^2$ in case of counterpropagating QH and QAH edge states. The presented toy model serves as a proof-of-principle and, in particular, the realistic scaling behavior can deviate from the analytic form shown in Eq.~\eqref{eqA:manyPuddles} . Deriving a microscopic model will be a subject of future work.
\clearpage

\end{document}